\documentclass[%
 reprint,
 amsmath,amssymb,
 aps,superscriptaddress
]{revtex4-1}

\usepackage{graphicx}
\usepackage{dcolumn}
\usepackage{bm}
\usepackage{lipsum}

\usepackage{color}

\newcommand{\beqar}{\begin{eqnarray}}
\newcommand{\eeqar}{\end{eqnarray}}
\newcommand{\bea}{\begin{eqnarray}}
\newcommand{\eea}{\end{eqnarray}}
\newcommand{\bcen}{\begin{center}}
\newcommand{\ecen}{\end{center}}
\newcommand{\ra}{\rightarrow}
\newcommand{\bra}[1]{\left< #1 \right|}
\newcommand{\ket}[1]{\left| #1 \right>}

\newcommand{\eps}{\varepsilon}
\newcommand{\lam}{\lambda}

\newcommand{\f}[2]{\frac{#1}{#2}}
\renewcommand{\b}[1]{\left({#1}\right)}
\renewcommand{\v}[1]{\vec{#1}}
\newcommand{\pd}[2]{\frac {\partial #1}{\partial #2}}

\renewcommand{\sb}[1]{\left[{#1}\right]}
\newcommand{\mean}[1]{\langle {#1} \rangle}
\newcommand{\infint}{\int_{-\infty}^{\infty}}

\newcommand{\tg}{\textcolor{green}}

\begin{document}
\title[Time Dependent Markovian Quantum Master Equation]{Time Dependent Markovian Quantum Master Equation}

\author{Roie Dann}
\email{roie.dann@mail.huji.ac.il}
\affiliation{The Institute of Chemistry, The Hebrew University of Jerusalem, Jerusalem 91904, Israel.}
\affiliation{The Kavli Institute for Theoretical Physics, The University of California, Santa Barbara, California, USA, CA 93106}

\author{Amikam Levy}
\affiliation{Department of Chemistry, University of California Berkeley, Berkeley, California 94720, USA}
\affiliation{The Sackler Center for Computational Molecular Science, Tel Aviv University, Tel Aviv 69978, Israel}

\author{Ronnie Kosloff}
\affiliation{The Institute of Chemistry, The Hebrew University of Jerusalem, Jerusalem 91904, Israel.}
\affiliation{The Kavli Institute for Theoretical Physics, The University of California, Santa Barbara, California, USA, CA 93106}

\email{roie.dann@mail.huji.ac.il}

\begin{abstract}
We construct a quantum Markovian Master equation for a driven system coupled to a thermal bath.
The derivation utilizes an explicit solution of the propagator of the driven system. This enables the validity of the Master equation to be extended beyond the adiabatic limit.
The Non-Adiabatic Master Equation (NAME) is derived employing the weak system-bath coupling limit.
The NAME is valid when a separation of timescales exists between the bath
dynamics and  the external driving.
In contrast to the adiabatic Master equation, the NAME leads to coupled equations of motion for the population and coherence. 
We employ the NAME to solve the example of an open driven time-dependent harmonic oscillator. For the harmonic oscillator the NAME predicts the emergence of coherence associated with the dissipation term. As a result of the non-adiabatic driving the thermalization rate is  suppressed.
The solution is compared with both numerical calculations and the adiabatic Master equation.
\end{abstract}
\maketitle

 \section{Introduction}
 \label{sec:intro}
All physical systems in nature, small or large, are affected  to some extent by an external environment. 
The theory of open quantum systems incorporates the influence of the environment on the dynamics of a quantum system in a concise manner. 
In this framework the aim is to find the reduced dynamical description of the primary system while tracing out the environment. 
The dynamical map describing the system's evolution is required to be completely positive and trace  preserving (CPTP). This mathematical property is required to allow a consistent physical interpretation of the quantum dynamics.  
The most general form of a CPTP dynamical reduced description of divisible maps is given by 
the Gorini-Kossakowski-Lindblad-Sudarshan (GKLS) Markovian master equation \cite{lindblad1975completely,lindblad1976generators,gorini1976completely}. 
There are several options for deriving the GKLS equations from first principles. 
In this study we will follow the path of the Born-Markov weak system bath coupling derivation originally derived by Davies \cite{davies1974markovian}. 

The GKLS form fulfills the thermodynamical requirements such as the first and second laws of thermodynamics. \cite{alicki2018introduction,kosloff2013quantum,marcantoni2017entropy,argentieri2015complete}.
This Master equation is a template in many fields, such as  in quantum optics \cite{carmichael2009open,scully1999quantum}, quantum measurement \cite{wiseman2009quantum}, quantum information \cite{nielsen2002quantum} and quantum thermodynamics \cite{kosloff2013quantum}. 

The original Davies construction assumes a static system Hamiltonian leading to a Master equation, 
where the environment is expressed through its second order correlation functions and bath modes matching the system's intrinsic frequencies. 
This Davis approach has been generalized for the dissipative dynamics of periodically driven systems using the Floquet theory \cite{geva1995relaxation,alicki2012periodically,LevyRefrigerators2012,
szczygielski2013markovian,kosloff2014quantum}, 
and adiabatic driving \cite{davies1978open,albash2012quantum,childs2001robustness,kamleitner2013secular,sarandy2004consistency}.
Recently, Yamaguchi et al., generalized the Master equation beyond the adiabatic regime \cite{yamaguchi2017markovian},  
where the final form  of the Master equation was  identical  to the adiabatic equation of Albash et al.  \cite{albash2012quantum}. 
In this paper we derive a Non-Adiabatic Master Equation (NAME) going beyond the approximations of Albash and Yamaguchi.

In the derivation of the NAME a Lie algebraic structure of the driven system evolution operators is employed.  
The outcome is a  time-dependent GKLS operator structure with time-dependent decay rates.  
Unlike the case of the adiabatic GKLS equation, the equations of motion of the population and coherence are coupled and this leads to generation of coherence associated the dissipative term.

One of the most well-studied examples of open quantum systems is the Master equation of the quantum harmonic oscillator. The same equation is employed in many physical disciplines such as quantum optics, ions in a Paul trap, opto-mechanical oscillators, and vibrational modes of molecules in solution.
We would like to extend such scenarios to processes with an explicit time dependent Hamiltonian.
A quantum harmonic oscillator with a varying frequency, coupled to a Bosonic bath, is employed to demonstrate the NAME. 
The results for this model predict non-vanishing coherence due to the inhomogeneous terms in the equations of motion. 
These terms define the {\em instantaneous attractor}
which provides a new insight of the relation between the system and bath for non-adiabatic processes.
The NAME construction enables a thermodynamically consistent study of driven systems coupled to the environment such
as isothermal strokes in a quantum Carnot engine \cite{geva1992quantum}, and quantum control of open systems \cite{koch2016controlling,levy2018noise,vinjanampathy2018speeding}.

We begin by presenting in section \ref{gen_deriv} a general derivation of the NAME, study the asymptotic limits of the equation (Sec. \ref{sec: Asymp}) and present an analysis of the approximations in Sec. \ref{Approx validity sec}. In section \ref{HO_expample_sec} 
we study a specific example of a driven harmonic oscillator and verify the validity of the NAME by numerical methods (Sec. \ref{Numerical_analysis_sec}). 
This paper is accompanied by  detailed appendices that include the explicit derivation of the NAME and the numerical simulation details.

\section{Derivation of the general Non-Adiabatic Master  Equation}
\label{gen_deriv}

The starting point of the derivation of the NAME is a system coupled to a single bath.
We assume that the  dynamics of the composite system  is  closed and follows a unitary evolution generated by the composite Hamiltonian \cite{von1955mathematical,breuer2002theory}
\begin{equation}
\hat H\b t=\hat H_{S}\b t+\hat H_{B}+\hat H_{I}~~.
\label{eq:ham1}
\end{equation}

In (\ref{eq:ham1}) $\hat H_{S}\b t$ and $\hat H_{B}$ are the system and bath Hamiltonians and $\hat H_{I}$ is the system-bath interaction term, which can be expressed as
\begin{equation}
\hat H_{I}=\sum_{k}g_{k}\hat A_{k}\otimes \hat B_{k}~~.
\label{eq:H_I}
\end{equation}
Here, $\hat A_{k}$ and $\hat B_{k}$ are the Hermitian operators of the system and bath, respectively, and $g_{k}$ are the coupling strength parameters. 
Following the standard perturbation  expansion, the first step is a transformation to the interaction picture with respect to the $\hat H_S \b t$ and bath Hamiltonians,
\begin{equation}
{\tilde{H}}\b t=\hat U_{S}^{\dagger}\b{t}\hat U_{B}^{\dagger}\b{t}\hat H\b t \hat U_{B}\b{t} \hat U_{S}\b{t} ~~~,
\end{equation}
where the free bath propagator is $\hat{U}_{B}\b{ t,0}\equiv \hat{U}_{B}\b t=e^{-i\hat H_{B}t/\hbar}$, and $\hat U_{S}\b{ t,0}\equiv\hat U_{S}\b t = {\cal{T}} \exp \b{-\f{i}{\hbar}\int_{0}^{t} \hat{H}_S \b{t'} dt'}$. Here, $\cal{T}$ is the time-ordering operator and the tilde symbol is assigned to operators in the interaction picture.
The system propagator $\hat U_{S}(t)$  is   the solution of the Schr\"odinger equation for a time-dependent Hamiltonian
\begin{equation}
i \hbar \pd{\hat U_S \b {t}}{t} =\hat H_S \b t \hat U_S \b {t} ~~,~~\hat U_S \b {0}=I~~~.
\end{equation}
In the interaction picture, the interaction Hamiltonian takes the form:
\begin{equation}
{\tilde{H}}\b t={\tilde{H}_I}\b t=\sum_{k}g_{k} {\tilde{A}}_{k}\b t\otimes  {\tilde{B}}_{k}\b t
\label{interaction}
\end{equation} 
where the interaction picture operators of the bath and system are $ {\tilde{B}}_{k}\b t=e^{i\hat H_{B}t/\hbar}\hat B_{k}e^{-i\hat H_{B}t/\hbar}$ and ${\tilde{A}}_{k}\b t=\hat U_{S}^{\dagger}\b{t}\hat A_{k}\hat U_{S}\b{t}$. 
\par
To obtain a Master equation of the GKLS form, the Liouville von Neumann equation is expanded up to second order in the coupling strength $g_k$, relying on the weak coupling limit. Furthermore, the Born-Markov approximation is employed involving three main assumptions, \cite{breuer2002theory}: 
\begin{enumerate}
\item{The quantum system and the bath are uncorrelated, such that $\hat \rho\b t=\hat \rho_{S}\b t\otimes\hat \rho_{B}$.}
\item{ The bath correlation functions decay much faster than the system's relaxation rate and internal dynamics.}
\item  {The state of the bath is assumed to be a thermal stationary state,\\  $\hat \rho_{B}=e^{-\beta \hat H_{B}}/\text{tr}\b{e^{-\beta \hat H_{B}}}$.}
\end{enumerate}
These assumptions with the second order perturbation theory  lead to  the Markovian quantum master equation 
\begin{multline}
\f d{dt} {\tilde{\rho}}_{S}\b t=\\-\f {1}{\hbar^2}\int_0^{\infty}{ds\text{tr}_B\sb{{\tilde{H}}\b t,\sb{ {\tilde{H}}\b{t-s}, {\tilde{\rho}}_{S}\b t\otimes {\tilde{\rho}}_{B}}}}~~~.
\label{quantum markovian master eq.}
\end{multline}
This equation has also been derived using the time  convolution-less technique \cite{shibata1977generalized,hashitsumae1977quantal}.

To reduce Eq. (\ref{quantum markovian master eq.}) from an integro-differential to a differential form we introduce the set of time-independent eigenoperators, $\{ {\tilde{F}} \}$, of the propagator  $\hat{U}_S \b t $. The eigenoperators are defined by the equation,
\begin{equation}
 \tilde{F}_j\b t=\hat{U}_S^\dagger \b t \hat{F}_{j}\b 0  \hat{U}_S \b t=\lam_j \b t \hat{F}_{j}\b 0 ~~,
 \label{F_def_eq}
\end{equation}
which is an eigenvalue equation in terms of operators, where $\lam_j \b t =|\lam_j \b t| e^{i\phi_{j}\b t}$ are complex eigenvalues. 
The unitarity  of $\hat{U}_S\b t$ and the algebraic properties guarantee the existence of operators $\tilde{F}_{j}$, see Appendix \ref{App:eigenoperators}.
 The set $\{ {\tilde{F}} \}$  form a complete basis of the system's Lie algebra, allowing to expand ${\tilde{A}}_{k}\b t$ in terms of the eigenoperator basis, 
\begin{equation}
 {\tilde{A}}_{k}\b t=\sum_{j}\xi_{j}^{k} \b t e^{i\theta_{j}^k\b t} {\hat{F}}_{j}\b 0
\label{Decomposition to eigen operators}
\end{equation}
where $\theta_j^k \b t$ includes the time-dependent phase $\phi_j$ of $\tilde{F}_j\b t$, and any phase associated with $\tilde{A}_k \b t$. Similarly $\xi_j^k$ is a function of $|\lam_j\b t|$ and any explicit time dependence of the system interaction operator.
The time-dependent coefficients satisfy $\xi_{j}^{k}\b t$, $\theta_{j}^k\b t\in\mathbb{R}$ and $\xi_{j}^{k}\b t>0$, see Appendix \ref{sec:expansion set}.
In the following we omit the time dependence of the eigenoperators at initial time, $\hat{F}_{j} \equiv \hat{F}_{j} \b 0$.


Inserting equation (\ref{Decomposition to eigen operators}) in equation (\ref{quantum markovian master eq.}) we obtain after some algebra
\begin{multline}
\f d{dt} {\tilde{\rho}}_{S}\b t=\f {1}{\hbar^2}\sum_{k,k',j,j'}\int_0^{\infty}{ds}\,\xi_{j'}^{k'}\b t\xi_{j}^{k}\b{t-s}\\\times e^{i\theta_{j'}^{k'}\b t}e^{i\theta_{j}^{k}\b{t-s}} g_{k'}
g_{k}  \Big({{\hat{F}}_{j}\tilde{\rho}_{S} \b t {\hat{F_{j'}}}-  {\hat{ F}}_{j'}{\hat{ F}}_{j} \tilde{\rho}_{S} \b t }\Big)\\\times  \text{tr}_{B}\{{\tilde B}_{k'}\b{t}{\tilde B}_{k}\b{t-s}\hat {\rho}_{B}\}+\text{h.c.}~~,
\label{eq_1}
\end{multline}
where h.c. denotes the hermitian conjugated expression.

Equation (\ref{eq_1}) describes dynamics influenced by the past history of the driving protocol, 
incorporated by $\xi_{j}^{k}\b {t-s}$ and $\theta_{j}^k \b{t-s}$. 
The analytical solution for such an integro-differential equation presents a challenge \cite{nakajima1958quantum,zwanzig1960ensemble,breuer2001time,chang1993non}, 
and is not guaranteed to be completely positive, therefore further approximations are required.
We assume that the bath dynamics is  fast compared to  the driving rate which determines the adiabatic parameter $\mu$.
In general the adiabatic parameter is defined as,
\begin{equation}
 \mu  = \text{max}_{t,k,l}\sb{\f{\bra {k\b t}{\dot{\hat{H}}_S\ket{l \b t}}}{|E_{k}\b t-E_{l}\b t|^{2}}} ~~,
\label{mu_eq}
\end{equation}

where $E_j(t)$ and $\ket{j(t)}$  are the instantaneous eigenenergies and eigenstates of the Hamiltonian $\hat H_S(t)$ \cite{mostafazadeh1997quantum}.
A slow change of the driving protocol relative to the bath dynamics will lead to a slow change of $\xi_{j}^{k}\b t$ and $\theta_{j}^k\b t$ relative to the bath decay rate. 
This translates to a relation between the typical timescales:  The bath correlation  decay time, $\tau_{B}$, should be much shorter than the non-adiabatic timescale, $\tau_d$, an additional timescale which emerges in the derivation, associated with the change in the driving protocol, Cf. Sec \ref{Approx validity sec}.   For $s\in\sb{0,\tau_{B}}$ and  $s\ll t$,  $\xi_{j}^{k}\b{t-s}$ can be approximated by a polynomial expansion in orders of $s$,
\begin{equation}
\xi_{j}^{k}\b{t-s}\approx\xi_{j}^{k}\b t-\f{d}{dt}\xi_{j}^{k}\b{t}s~~.
\label{eq:xiexp}
\end{equation}
The amplitudes are influenced directly by the driving, hence, it can be assumed that
in the regime $s\sim \tau_B$ and $s<\tau_B$ (for slow change in the driving) the second term on the RHS is negligible relative to the amplitude $\xi_{j}^{k}\b t$, 
obtaining  $\xi_{j}^{k}\b{t-s}\approx\xi_{j}^{k}\b t$. It is possible also to include the first order terms in $s$, leading to a small correction of the decay rates (see Appendix \ref{sec:higher_order} on higher order corrections). 
 
For $s>\tau_{B}$ the bath correlation functions decay rapidly, therefore the contribution to the integral can be neglected. 

A similar approximation is performed for the phases by expanding $\theta_{j}^k\b{t-s}$ around $t$ up to first order, this order is the dominant contribution to the dynamics, hence it included in the derivation.
\begin{equation}
\theta_{j}^k\b{t-s}\approx\theta_{j}^k\b t-\f{d}{dt}\theta_{j}^k\b ts=\theta_{j}^k\b t+\alpha_{j}^k\b ts~~,
\label{theta taylor exp}
\end{equation}
where the second term in the expansion is defined as $\alpha_{j}^{k}\b t\equiv-\f d{dt}\theta_{j}^{k}\b{t-s}|_{s=0}$.
Inserting the expansions, Eq. (\ref{theta taylor exp}), into Eq. (\ref{eq_1}) leads to:
\begin{multline}
\f d{dt}\tilde{\rho}_{S}\b t=\sum_{k,k',j,j'}g_{k}g_{k'}\xi_{j}^{k}\b t\xi_{j'}^{k'}\b te^{i\theta_{j'}^{k'}\b t}e^{i\theta_{j}^k\b t}\\\times
\Gamma\b{\alpha_{j}^k\b t}\Big({ {\hat{F}}_{j}\tilde{\rho}_{S}\b t {\hat{F_{j'}}}-   {\hat{ F}}_{j'} {\hat{ F}}_{j}{\tilde{\rho}}_{S} \b t  }\Big)+\text{h.c.}
\label{beforeRWA}
\end{multline}
where the Fourier transform of the instantaneous bath correlation function  is given by 
\begin{equation}
\Gamma_{kk'}\b{\alpha_{j}^k\b t}=\f {1}{\hbar^2}\int_0^{\infty}{ds}e^{i\alpha_{j}^k\b ts}\text{tr}_{B}\{\hat{B}_{k'}\b{t}\hat{B}_{k}\b {t-s}\hat{\rho}_{B}\}~~.
\label{bath_correlations_Fourier}
\end{equation}
To simplify, we decompose $\Gamma$ to a real and pure imaginary part,
\begin{equation}
\Gamma_{kk'}\b{\alpha}=\f 12\gamma_{kk'}\b{\alpha}+iS_{kk'}\b{\alpha}.
\end{equation}
Here, $\gamma_{kk'}\b{\alpha}$ can be written as $\gamma_{kk'}\b{\alpha}=\f {1}{\hbar^2}\infint ds e^{i\alpha s}\mean{\hat{B}_{k}\b s \hat{B}_{k'}\b 0\rho_{B}}_{B}$, where $S_{kk'}\b{\alpha}=\f 1{2i}\b{\Gamma_{kk'}\b{\alpha}-\Gamma_{k'k}^{*}\b{\alpha}}$, and $\langle~~\rangle_B$ is the average over the bath's thermal state.

In order to obtain a Master equation in the GKLS form the secular approximation is  required. 
The approximation neglects fast oscillating terms in the Master equation, which average to zero in the time resolution of interest. In such a regime, assuming no degeneracy in the Bohr frequencies, the terms for which $\theta_{j'}^{k'}\b t\neq-\theta_{j}^k\b t$ 
oscillate rapidly relative to the relaxation dynamics and averages to zero.

Performing the secular approximation leads to the non-adiabatic-master-equation (NAME) in the interaction representation:
\begin{multline}
\f d{dt}\tilde{\rho}_{S}\b t=-\f{i}{\hbar}\sb{\tilde{H}_{LS}\b t,\tilde{\rho}_{S}\b t}\\
+\sum_{k,j}~\b{\xi_j^k\b t}^2  g_{k}^2\gamma_{kk}\b{\alpha_{j}^k\b t}\\
\b{\hat{F}_{j}\tilde{\rho}_S\b t \hat{F}_{j}^{ \dagger}-\f 12\{\hat{F}_{j}^{ \dagger}\hat{F}_{j} ,\tilde{\rho}_{S}\b t\}}~~~.
\label{non_adiabatic_master_eq}
\end{multline}


Here, $\tilde{H}_{LS} \b t$ is the time-dependent Lamb-type shift Hamiltonian in the interaction representation, $\tilde{H}_{LS}\b{t}= \sum_{k,j}\hbar S_{kk}\b {\alpha_{j}^k \b t} \hat{F}_{j}^{\dagger} \hat{F}_{j}$,  ($\hat{F}_{j} \equiv \hat{F}_{j} \b 0$).

The decay rates in (\ref{non_adiabatic_master_eq}) are all positive, hence, the equation has a GKLS form, guaranteeing a CPTP map for the system's state.
Equation (\ref{non_adiabatic_master_eq}) has a very similar form to the Quantum Markovian Adiabatic equation of Albash \cite{albash2012quantum}
and the generalization of Yamaguchi \cite{yamaguchi2017markovian}.  
The differences which arise are the scalar rate coefficients and the dissipative generator operators $\hat F_j$. 
As will be shown in the next sections, these differences result in different qualitative and quantitative behavior.

\section{Asymptotic limits of the NAME}
\label{sec: Asymp}
The stationary Master equation as well as the adiabatic and periodically driven Master equations are asymptotic limits of the NAME
(\ref{non_adiabatic_master_eq}). 

\subsection{Periodic driving}  
The structure of the NAME, Eq.(\ref{non_adiabatic_master_eq}),  also holds when the system is driven by a periodic external field, see \cite{alicki2006internal,LevyRefrigerators2012}. The decomposition now reads
\begin{equation}
\tilde{A}_k(t) = \sum_j \xi_j^k e^{i\theta_j^k(t)}\hat{F}_j,
\end{equation}
where $\xi_j^k$ is time independent and $\theta_j^k\b t = (\omega_j + m\Omega)t$. The quasi-Bohar frequencies $\omega_j$ are the Floquet modes, $\Omega=2\pi/\tau $ with a period time $\tau$, and $m=0,\pm 1, \pm 2,...$. 
In this case, the operator $\tilde{F}_j$ is the part of $\tilde{A}_k(t)$ that rotates with frequency $\omega_j + m\Omega$, and 
the summation  in Eq.(15) is replaced by $\sum_j \rightarrow \sum_{m\in \mathbb{Z}} \sum_{\{\omega_j\}}$.

\subsection{Adiabatic limit}

A quantum adiabatic process is such that an initial energy state, $\ket {\eps_a \b 0}$, follows the corresponding time-dependent  eigenstate,  $\ket {\eps_a \b t}$, of the instantaneous Hamiltonian, $\hat H_S \b t$,
$$\hat H_S \b t \ket{\eps_a \b t} = \eps_a \b t \ket{\eps_a \b t}~~~.$$
Following the derivation in \cite{albash2012quantum}, in the adiabatic limit, the propagator can be represented in terms of the instantaneous energy eigenstates,
\begin{equation}
\hat U_{S}\b{t,t'}\approx \hat U_{S}^{\text{adi}}\b{t,t'}=\sum_{a}\ket{\eps\b t}\bra{\eps\b{t'}}e^{-i\lam_{a}\b{t,t'}}~~~.
\end{equation}
The phase is given by $\lam_a \b {t,t'} = \hbar^{-1}\int_{t'}^{t} d\tau \sb{\eps_a \b \tau -\phi_a \b \tau}$, where $\{ \eps_a \b{t} \}$ are the instantaneous energies and $\phi_a \b t = i\mean{\eps_{a}\b t |\dot{\eps}_{a}\b t}$ is the Berry phase \cite{berry1984quantal,berry1987quantum}.
 
The system operators in the interaction picture are calculated using $U_{S}^{\text{adi}}\b{t,t'}$:
\begin{multline}
\tilde{A}_{k}\b t=U_{S}^{\text{adi}\dagger}\b{t,0}\hat{A}_{k}U_{S}^{\text{adi}}\b{t,0}\\
=\sum_{a,b}\bra{\eps_{a}\b t}\hat{A}_{k}\ket{\eps_{b}\b t}e^{-i\lam_{ba}\b{t,0}}\ket{\eps_{a}\b 0}\bra{\eps_{b}\b 0}~~~.
\label{eq:A adiabat}
\end{multline}  

We identify the expansion set operators as $\hat{F}_{ba}=\ket{\eps_{a}\b 0}\bra{\eps_{b}\b 0}$, the amplitude by $\xi_{ba}^{k}\b t=\bra{\eps_{a}\b t}\hat{A}_{k}\ket{\eps_{b}\b t}$, and the phases as:
\begin{multline}
\theta_{ba}\b{t,t'}=\lam_{ba}\b{t,t'}\equiv \lam_{b}\b{t,t'}-\lambda_{a}\b{t,t'} \\
=\f{1}{\hbar} \int_{t'}^{t}{d\tau\sb{\b{\eps_{b}\b{\tau}-\eps_{a}\b{\tau}}
-\b{\phi_{b}\b{\tau}-\phi_{a}\b{\tau}}}}~~~.
\end{multline}
Here, the indices  $b,a$ can be replaced by a single index $j$, reconstructing Eq. (\ref{Decomposition to eigen operators}).
Similarly to the derivation in section \ref{gen_deriv}, we expand the phase, $\theta_{ba}\b {t-s,0}$ at the vicinity of  $t$. The first order term becomes
\begin{multline}
\theta_{ba}\b{t-s,0}\approx\theta_{ba}\b {t,0}-\f {d}{dt}\theta_{ba}\b {t,0} s\\
= \theta_{ba}\b {t,0} -\omega_{ba} \b t s +
\b{\phi_{b}\b{t}-\phi_{a}\b{t}}s~~,
\end{multline}
where $\omega_{ba} \b t = \b{\eps_{b} \b t - \eps_{a} \b t}/\hbar$ are the instantaneous Bohr frequencies. 
The third term on the RHS  is  first order in the adiabatic parameter $\mu$ \eqref{mu_eq}. The frequency $\phi$ is proportional to $\mu$, therefore in the adiabatic limit when $\mu \ll 1$, $\phi$ can be neglected. The frequency $\alpha_{ba} \b t$  becomes in this limit
\begin{equation}
\alpha_{ba}= \omega_{ba} \b t ~~~.
\label{eq:Adi_alpha}
\end{equation}
 
Inserting Eq. \eqref{eq:A adiabat} and \eqref{eq:Adi_alpha} into Eq. \eqref{non_adiabatic_master_eq} we obtain the Quantum Adiabatic Master equation, Eq. (54) in \cite{albash2012quantum}.
The static Master equation can be obtained for a time-independent Hamiltonian, $\hat{H}_S\b t = \hat{H}_S\b 0$.

\section{The NAME for the Driven Harmonic Oscillator}
\label{HO_expample_sec}

Next, we study the validity of the NAME for the driven harmonic oscillator coupled to a Bosonic bath. This model is relevant for a wide range of applications, including atomic, molecular and optical physics \cite{Rossnagel325,blickle2012realization}.
Here we employ the properties and structure of the SU(1,1) Lie algebra \cite{hirayama1993oscillator,yukawa1964internal} to derive the NAME.

The system is represented by the Hamiltonian
\begin{equation}
\hat{H}_{S}=\f{\hat{P}^{2}}{2m}+\f 12m\omega^{2}\b t\hat{Q}^{2} ~~~,
\label{sys_Ham}
\end{equation}
where $\hat Q$ and  $\hat P$ are the position and momentum operators, $m$ and $\omega\b t$ are the mass and frequency of the system. 
Closed form solutions of the free evolution of the second order operators has been obtained for a
constant adiabatic parameter, $\mu=\f{\dot{\omega}}{\omega^{2}}=const$ \cite{kosloff2017quantum}, Appendix \ref{Adiabatic propagation}.\\
In this case, the driving protocol of a time duration $t_f$ between frequencies $\omega \b 0$ and $\omega \b{t_f}$  is given by, 
\begin{equation}
\omega\b t=\f{\omega\b 0}{{1-\mu\omega\b 0t}}~~.
\label{eq:mu}
\end{equation}
The adiabatic parameter $\mu$ is uniquely determined by $\omega\b 0$, $\omega \b{t_f}$ and $t_f$, obtaining finite values for bounded frequencies.
The evolution of the isolated system is presented in Appendix \ref{Appendix:exp}, and is used to expand the interaction term, \eqref{interaction}, in terms of the eigenoperators \eqref{F_def_eq}.

\subsection{Coupling to the bath}
The harmonic oscillator is coupled linearly to a Bosonic thermal bath,
\begin{equation}
\hat{H}_{I}=\hat{Q} \otimes g\sum \hat{p}_{k}=i g\sum_{k}\sqrt{\f {m\omega_{k}}{2}}\hat{Q}\otimes \b{\hat{b}_{k}^{\dagger}-\hat{b}_{k}}~~,
\label{eq:Intterm}
\end{equation}
where $p_{k}$ is the $k$-th oscillator momentum operator and $\hat{b}_{k}$, $\hat{b}_{k}^{\dagger}$ are the corresponding annihilation and creation operators. Other choices of linear system-bath coupling are possible as in Ref. \cite{dietrich2018probing}. 

Following the derivation described in Section  \ref{gen_deriv}, $\tilde{Q}\b t$ is decomposed into the set of eigenoperators (see Appendix \ref{Appendix:exp}): 

\begin{equation}
\tilde{Q}\b t=\xi\b t\sum_{j=\pm}\hat{F}_{j}e^{i\theta_{j}\b t}~~,
\label{Q_eq}
\end{equation}
where $\hat{F}_j\equiv \hat{F}_j \b{0}= \tilde{F}_j \b{0}$.\\ 
The set of eigenoperators are a linear combination of the position and momentum operators
\begin{equation}
\hat{F}_{+} \b{t}=A\hat{Q}\b t+B\hat{P}\b t=\hat{F}_{-}^{\dagger}\b{t}~,
\label{eq:foperator}
\end{equation}
where $A=\f 12\b{i\f{\mu}{\kappa}+1}$ and $B=i\frac{1}{m\omega \b 0\kappa}$.
The amplitude is given by $\xi\b t=\sqrt{1-\mu\omega\b 0t}$ and the phases: 
\begin{equation}
\theta_{\pm}\b t=\mp\frac{\kappa}{2}\int_0^t{\omega\b{t'}dt'}=\pm\frac{\kappa}{2\mu}\text{log}\b{\f{\omega \b t}{\omega\b 0}}~~,
\label{theta_pm}
\end{equation}
where $\kappa=\sqrt{4-\mu^2}$. Notice that $\b{1-\mu\omega \b{0}t}$ is necessarily positive for physical $\omega\b t$, Eq. \eqref{eq:mu}, leading to a real value for the accumulated phases. 

In order to perform the secular approximation we analyze the time dependence of $\theta_{\pm}(t)$. The approximation is valid when $|2\theta_{+}\b t|$ oscillates rapidly relative to the decay frequency, $\tau_R^{-1}$. This adds a restriction on the range of $\theta_{\pm}\b t$ and $\omega\b t$ with respect to the driving protocol, leading to the inequality $|\theta_{\pm}\b {t}|\gg 1$ for $t<\tau_R$. A full analysis of the approximation and regime of validity are presented  in Sec. \ref{Approx validity sec}.


Following the general derivation for a specific $\xi \b t$, $\hat{F}_j$ and $\theta_j$ the  correlations one-sided Fourier transforms, $\Gamma_{kk'}$ in Eq. (\ref{bath_correlations_Fourier}), can be calculated, determining the dissipative rates in the NAME, Eq. (\ref{non_adiabatic_master_eq}). 

By collecting equations (\ref{theta_pm}), (\ref{Q_eq}) and (\ref{non_adiabatic_master_eq}) the NAME in the interaction representation becomes:
\begin{multline}
\f d{dt}\tilde{\rho}_{S}\b t=-\f{i}{\hbar}\sb{\tilde{H}_{LS}\b t,\tilde{\rho}_{S}\b t}+|\xi\b t|^{2}\gamma\b{\alpha\b t}\\\times
\Bigg(\hat{F}_{+}\hat{\rho}_{S}\b t\hat{F}_{-}-\f 12\{\hat{F}_{-}\hat{F}_{+},\tilde{\rho}_{S}\b t\}\\
+e^{-\hbar \alpha\b t /k_B T}
\b{\hat{F}_{-}\tilde{\rho}_S\b t \hat{F}_{+}-\f 12\{\hat{F}_{+}\hat{F}_{-},\tilde{\rho}_{S}\b t\}} \Bigg)~,
\label{non_adi_HO}
\end{multline}
where $k_B$ is the Boltzmann constant, $T$ is the bath temperature, $\alpha \b t=\f{\kappa}{2} \omega \b t$ and $\hat{F}_+\equiv\hat{F}_+\b 0$. 
The time-dependent rate coefficient has the form, 
\begin{equation}
\gamma\b{{\alpha}\b t}=\f{m \pi}{ \hbar}{{\alpha}\b t} J\b{{\alpha}\b t}\b{\bar{N}\b{{\alpha}\b t}+1}
\end{equation}
where $J \b{\alpha}$ is the spectral density function determined by the density of bath states $f\b{\alpha}$  and the coupling strength $\chi\b{\alpha}$, $J\b{\alpha} = f\b{\alpha} \chi\b{\alpha}$ \cite{carmichael2009open} (Cf. Appendix \ref{sec:higher_order}). The factors $\bar{N}\b{\alpha}$ is the mean occupation number given by the Bose-Einstein statistics and $e^{-\hbar \alpha\b t /k_B T}$ is the instantaneous Boltzmann factor related to the effective time-dependent frequency $\alpha\b t$.

\subsection{Solution for the NAME}
\label{subsec:Solution_NAME}

For a time-independent problem it is convenient to transform to the Heisenberg picture, and obtain a set of coupled linear differential equations for the operators\tg{,} \cite{geva1995relaxation,kosloff2017quantum}. For Hilbert space of dimension $N$ one obtains $N^2-1$ equations which can be solved analytically or by standard numeric methods \cite{schaefer2017semi}.
In contrast, the solution is more complicated when the GKLS equation  has  an explicit time dependence.  
For such a case the solution for the operator $\hat{O}$ is given by \cite{breuer2002theory}:
\begin{equation}
\f{d}{dt} \hat{O}_H \b t = V^{\dagger}\b {t,0}\{ {\cal{L}}^{\dagger}\b t \hat{O}_H \b t \}~~,
\label{eq:A_Heis2}
 \end{equation} 
 subscript $H$ designates operators in the Heisenberg picture.
 The adjoint propagator takes the form:
\begin{equation}
 V^{\dagger} \b{t,t_0}= {\cal{T}}_{\ra}\exp{\int_{t_0}^{t} ds {\cal{L}}^{\dagger}\b s}~~,
 \label{eq:V_dag2}
\end{equation}
where ${\cal{T}}_{\ra}$ is the anti-chronological time-ordering operator and $V^{\dagger} \b{t,t_0}$ 
satisfies the differential equation
 $\pd{}{t} V^{\dagger}\b{t,t_0}=V^{\dagger}\b{t,t_0} {\cal{L}}^\dagger \b t$.
In order to obtain an equation of motion for $\hat{O}_H\b t$ (Eq. \ref{eq:A_Heis2}), one first needs to apply the time-dependent adjoint generator at time $t$ on the operator at initial time, and then propagate the solution in time with  $V^{\dagger}\b {t,0}$. In general, this proves to be difficult as a result of non-commutativity of ${\cal{L}} ^\dagger \b s$ at different times, requiring time-ordering in Eq. (\ref{eq:V_dag2}). To circumvent the problem of time-ordering in the Heisenberg representation, we solve the dynamics of the density matrix.

Solving the NAME in the interaction picture simplifies the analysis. The equation is expressed in terms of normalized creation and annihilation operators: $\hat{b}=\sqrt{c}\hat{F}_{+}$ and $\hat{b}^{\dagger}=\sqrt{c}\hat{F}_{-}$, where $c = \b{2\hbar \text{Im}\b{A^* B}}^{-1}$ for $A$ and $B$ introduced in Eq. (\ref{eq:foperator}) ($\hat{b}\equiv\hat{b}\b 0$). 
These operators satisfy the Bosonic annihilation and creation commutation relation $\sb{\hat{b},\hat{b}^{\dagger}}=1$ , allowing to cast the NAME in  the simple form. Assuming that the Lamb-shift term is negligible, we obtain,
\begin{multline}
\f d{dt}\tilde{\rho}_S \b t=k_{\downarrow}\b t \b{\hat{b}\tilde{\rho}_S \b t\hat{b}^{\dagger}-\f 12\{\hat{b}^{\dagger}\hat{b},\tilde{\rho}_S \b t \}}\\+k_{\uparrow} \b t\b{\hat{b}^{\dagger}\tilde{\rho}_S \b t\hat{b}-\f 12\{\hat{b}\hat{b}^{\dagger},\tilde{\rho}_S \b t\}}~~,
\label{eq:NAMEOH}
\end{multline}
where $k_{\downarrow}\b t=\f{m \pi c}{ \hbar}{{\alpha}\b t}J\b{{\alpha}\b t}\b{\bar N\b{{\alpha}\b t}+1}$ and $k_{\uparrow} \b t=\f{m \pi c}{ \hbar}{{\alpha}\b t}J\b{{\alpha}\b t}\b{\bar{N}\b{{\alpha}\b t}}$.

We assume an initial squeezed Gaussian state in terms of the operator basis $\{{\hat{b}}^{\dagger} \hat{b},\hat{b}^2,\hat{b}^{\dagger 2},\hat{I}\}$, which is preserved under the dynamics of the NAME, \cite{rezek2006irreversible}:
\begin{equation}
\tilde{\rho}_S \b t =\f{1}{Z\b t} e^{\gamma \b t \hat{b}^2}e^{\beta \b t  \hat{b}^{\dagger}\hat{b}}e^{\gamma^* \b t \hat{b}^{\dagger 2}}
\label{eq:rho_S}
\end{equation}
where $Z\b t$ is the partition function:
\begin{equation}
Z\b t \equiv Z\b{\beta,\gamma}=\f{e^{-\beta}}{\b{e^{-\beta}-1}\sqrt{1-\f{4|\gamma|^{2}}{\b{e^{-\beta}-1}^{2}}}}~~~.
\end{equation}

For the general case of a finite Lie algebra $\tilde{\rho}_S \b t$ can be expressed in terms of a generalized Gibbs state (ensemble) density operator \cite{langen2015experimental,jaynes1957information}, the squeezed Gaussian is a special case of such a state, see Appendix \ref{app:GGS}.

Inserting Eq. (\ref{eq:rho_S}) into Eq. (\ref{eq:NAMEOH}) and multiplying the equation of motion by $\tilde{\rho}_S^{-1}$ leads to $\b{\f{d}{dt}\tilde{\rho}_S} \tilde{\rho}_S^{-1}=\b{{\cal{L}}\tilde{\rho}_S}\tilde{\rho}_S^{-1}$ where $\cal{L}$  is the generator in the RHS of Eq. (\ref{eq:NAMEOH}). Utilizing the Baker-Housdorff relation the RHS is decomposed to a linear combination of the algebra operators. Comparing both sides of the equation, term by term, we obtain two coupled differential equations for $\gamma\b t$ and $\beta\b t$, (a detailed derivation appears in the Appendix \ref{sec:HLCexp}):
\begin{gather}
\label{eq:betagamma}
\dot{\beta}=k_{\downarrow}\b{e^{\beta}-1}+k_{\uparrow}\b{e^{-\beta}-1+4e^{\beta}|\gamma|^{2}}
\\
\dot{\gamma}=\b{{k}_{\downarrow}+{k}_{\uparrow}}\gamma-2{k}_{\uparrow}\gamma e^{-\beta}~~~.
\nonumber
\end{gather}
Notice that the rates $k_{\downarrow}$ and $k_{\uparrow}$ are in general time-dependent, increasing the difficulty for obtaining an analytical solution.
Once $\beta\b t$ and $\gamma \b t $ are obtained the expectation values of the set of operators can be retrieved from Eq. (\ref{eq:rho_S}), thus, circumventing 
the use of the Heisenberg representation.
Eq. (\ref{eq:betagamma}) was solved numerically using the Runge-Kutta-Fehlberg method and the solutions of $\beta \b t$ and $\gamma \b t$ are utilized to calculate expectation values, see Appendix \ref{sec:HLCexp}.


In order to analyze the system dynamics we define two additional time-dependent  operators:
\begin{equation}
\hat{L} \b t  = \f {\hat P^2 }{2m} - {\f{1}{2} m \omega \b t \hat Q^2} 
\,\,\,\,\text{and}\,\,\,\,
\hat{C} \b t  = \f{\omega \b t}{2}\b{\hat Q \hat P +\hat P \hat Q}~~.
\label{eq:L}
\end{equation}
 
 \begin{figure}[htb!]
\centering
\includegraphics[scale=0.27]{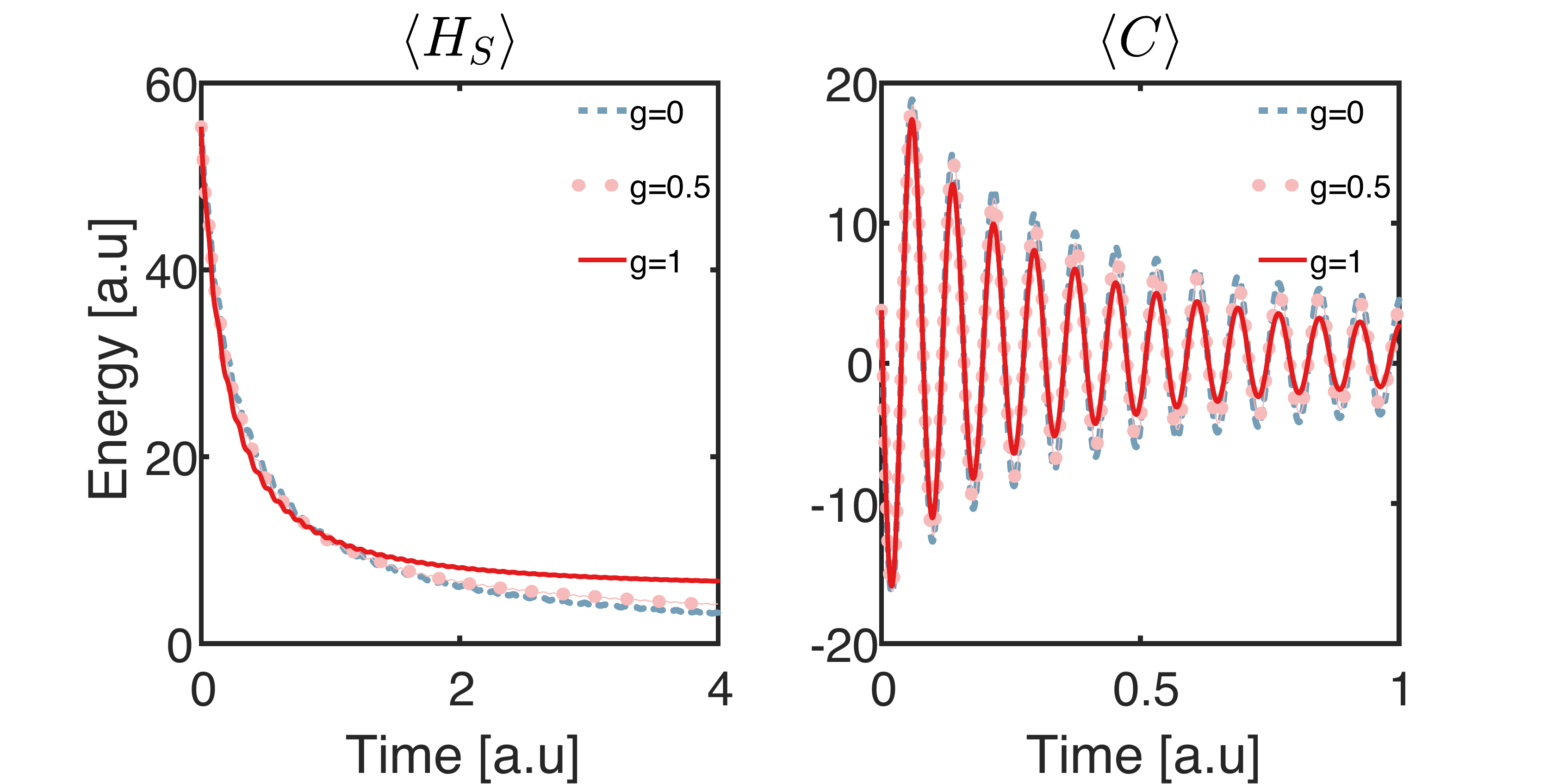} 
\caption{\label{figureone} 
System dynamics as generated from the NAME for different coupling strengths $g$ ($g=0$ represents isolated dynamics). The left panel shows the expectation value of the energy as a function of time and the right panel shows the position momentum correlation, $\mean{\hat{C}}$ as a function of time. 
The chosen parameters are: $\mu=-0.1$, $\omega \b 0 =40$ and $T = 20$ where the initial conditions are $\beta\b 0 = -1$ and $\gamma \b 0 = 0.5$. This corresponds to an initial state described by $\mean{\hat H_S \b 0} \approx 55$, $\mean{\hat L \b 0}\approx -20.5$ and $\mean{\hat{C} \b 0}\approx 3.7 $. }
\end{figure}

The operators $\hat{L} \b t$ and $\hat{C} \b t$ together with $\hat{H}_S \b t$ and the identity  constitute a closed Lie algebra. These three operators define the state of the system, Appendix \ref{sec:HLCexp} \cite{kosloff2017quantum}. ${\hat{L}}$ is the difference between kinetic and potential energy and ${\hat C}$ is the position-momentum correlation, defining the squeezing of the state. Both expectation values vanish at thermal equilibrium. Since $\hat{L} \b t$ and $\hat{C}\b t$ do not commute with $\hat H_S\b t$ they can be employed to define
the coherence: ${\cal C} \equiv \f{\sqrt{\langle \hat{L}\rangle^2+\langle\hat{C}\rangle^2}}{\hbar \omega \b t}$ \cite{kosloff2017quantum}.
These operators can describe all thermodynamical equilibrium and out of equilibrium properties and are employed to reconstruct the
generalized Gibbs  state $\tilde \rho_S \b t$  \cite{kosloff2017quantum}. 



 \begin{figure}[htb!]
\centering
\includegraphics[scale=0.27]{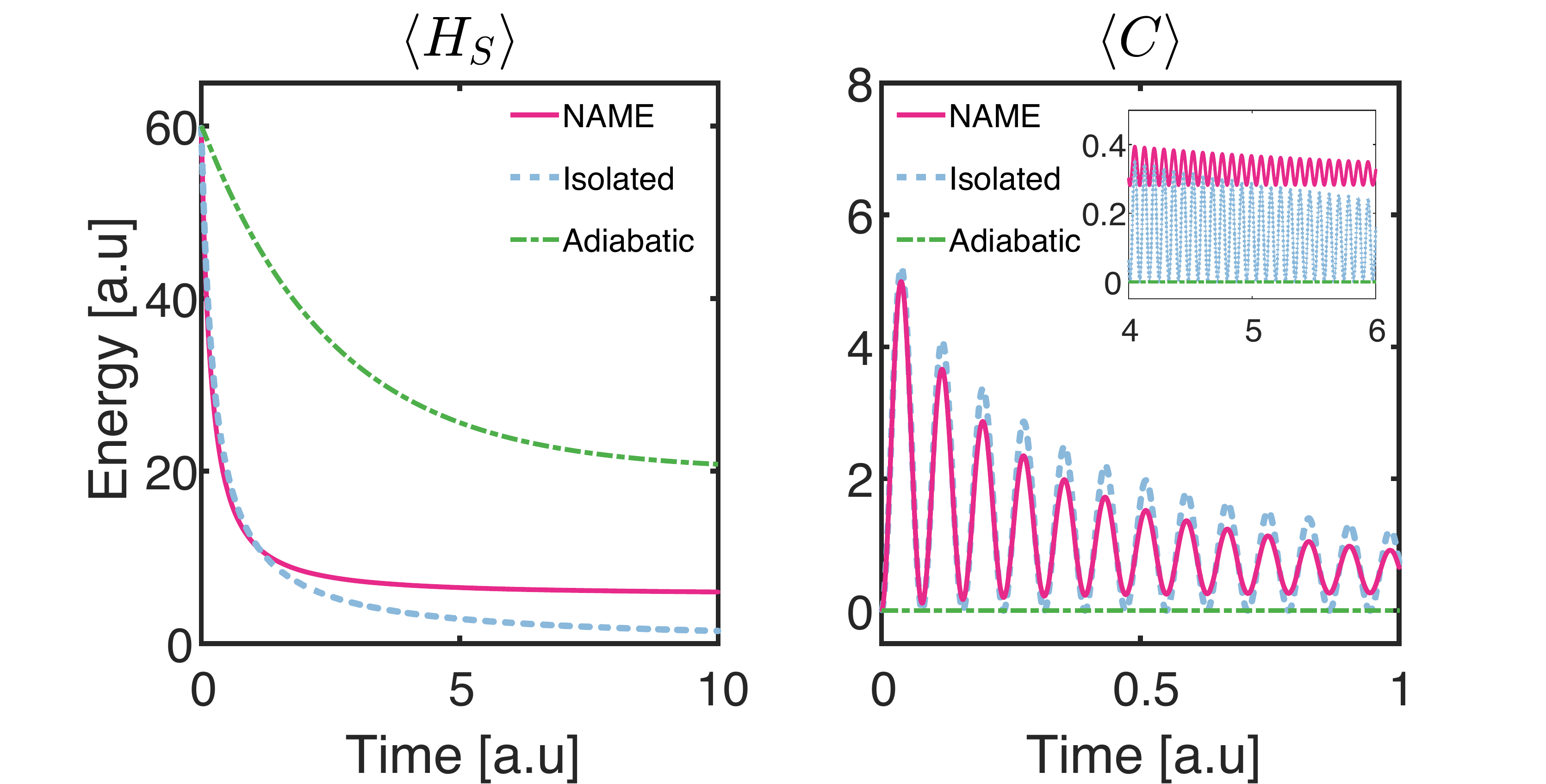} 
\caption{\label{figuretwo} 
The  system dynamics  generated from the NAME (pink, dark gray) is compared to the isolated quantum system (blue, dashed light gray)  and the {\em instantaneous attractor} (fixed point) of the adiabatic solution (green, dashed gray) for a parametric harmonic oscillator. 
The dynamics are represented by the system variables $\mean{\hat H_S\b{t}}$, $\mean{\hat L(t)}$ and $\mean{\hat{C}(t)}$.
Here, the chosen parameters are: $\mu=-0.1$, $\omega \b 0 =40$, $T = 20$ and $g=1$ where the initial conditions include no coherence $\mean{\hat{H}_S}=60$, $\mean{\hat{L}\b 0}= \mean{\hat{C}\b 0}=0$. }
\end{figure}

Using the formulation above, the expectation values of the operators $\mean{\hat H_S\b{t}}$, $\mean{\hat L(t)}$ and $\mean{\hat{C}(t)}$ are solved as a function of time.
Fig. \ref{figureone} shows a comparison between the solution of the NAME for different system-bath coupling strengths. 
The vanishing system-bath coupling term $g=0$ corresponds to the isolated case.
For $\mu<0$ the oscillator frequency decreases with time leading to a reduction of the system's energy  as seen in Appendix \ref{Adiabatic propagation}. 
The expectation value of $\mean{\hat{C}(t)}$ shows damped oscillations, similarly $\mean{\hat{L}(t)}$ oscillates with an opposite phase difference.   
These oscillations arise due to coupling between population and coherence, Eq. (\ref{eq:Adiprop}).  
 When $g >0$ the system energy increases due to energy flow from the bath. The observables  $\mean{\hat{L}(t)}$ and  $\mean{\hat{C}(t)}$ are suppressed at short time. At later times  $\mean{\hat{L}(t)}$ and  $\mean{\hat{C}(t)}$  increase with the coupling strength $g$, beyond the isolated case (see inset of Fig. \ref{figuretwo}).

\begin{figure}[htb!]
\centering{
\includegraphics[scale=0.27]{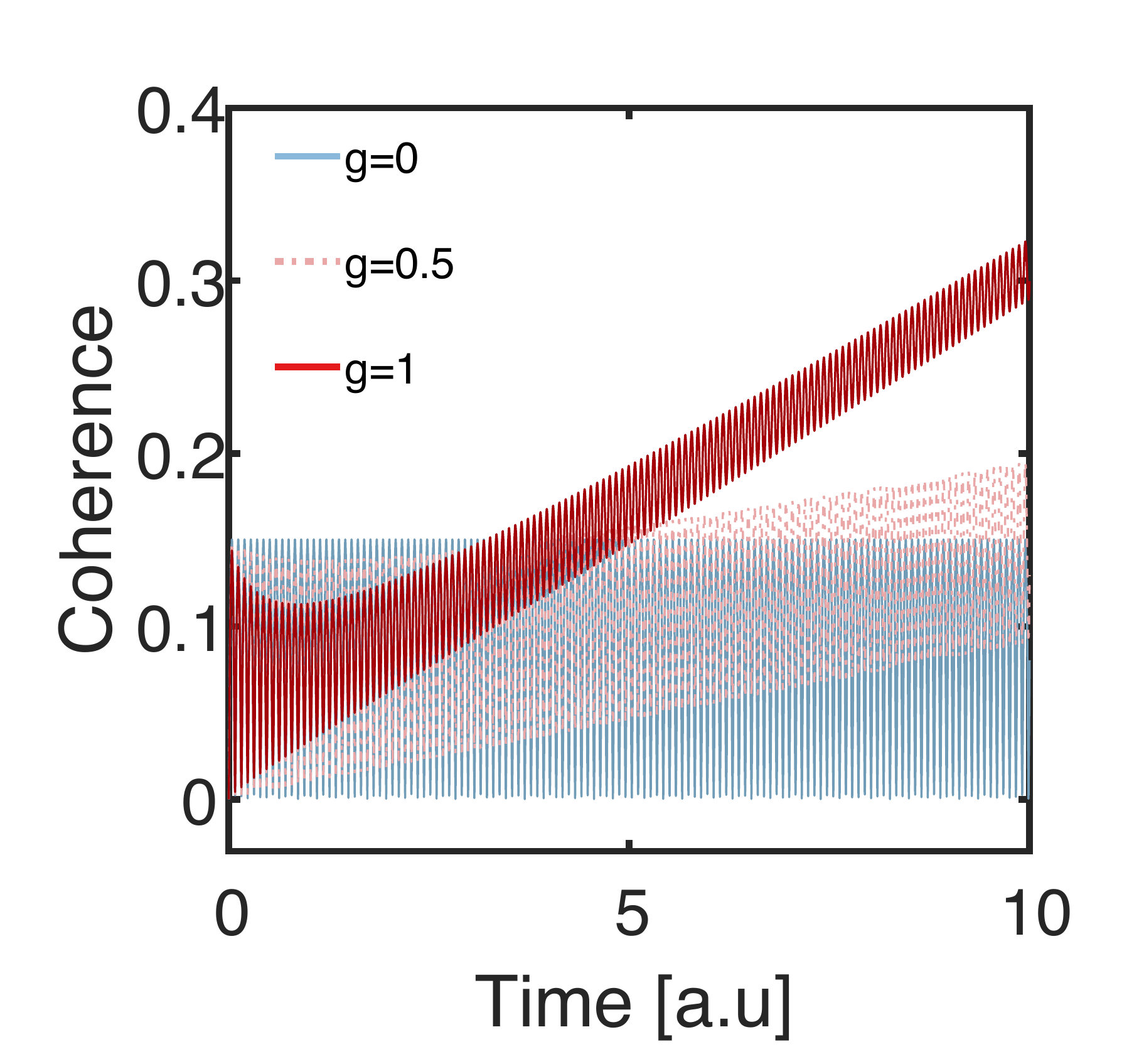}} 
\caption{\label{figurecoh}  
The dynamics of the coherence, ${\cal{C}}\equiv \f{\sqrt{\mean{\hat{L}}^2+\mean{\hat{C}}^2}}{\hbar \omega \b t}$, is presented for different system-bath coupling strengths. Increasing the system-bath coupling induces  an increase in coherence, associated with the non-adiabatic driving.
Model parameters and initial conditions are identical to Fig. \ref{figuretwo}. }
\end{figure}

Fig. \ref{figuretwo} shows the dynamics for an initial state which is diagonal in the energy eigenbasis ($\mean{\hat{L}\b 0}=\mean{\hat{C}\b 0}=0$). 
The analytical result of the NAME is compared to the isolated dynamics and the adiabatic Master equation. 
In the adiabatic case the system remains  diagonal in the energy eigenbasis at all times, with no generation of coherence throughout the dynamics. 
While non-adiabatic dynamics display a rise in coherence which oscillates in time.  The driving dresses the  system's state, leading to a rise in coherence attributed to both the unitary dynamics as well as to the dissipative term. At short times $\mean{\hat{L}(t)}$ and $\mean{\hat{C}(t)}$  are suppressed by the system-bath interaction as seen in Figure \ref{figuretwo}. However, at long times for non-adiabatic driving, $\mean{\hat{C}(t)}$ and $\mean{\hat{L}(t)}$ converge to a non-zero value. This is demonstrated in Figure \ref{figurecoh}, where we plot the dynamics of the coherence.
 
Figure \ref{figurecoh} shows the increase of coherence at later times for increasing bath coupling. The state of the system is mapped towards a direction which deviates from a direction defined by the instantaneous energy. This deviation can be understood from the structure of the jump operators $\hat{F}_{\pm}\b t$.
The non-adiabatic driving modifies the jump operators, which differ from the instantaneous (adiabatic) jump-operators, $\hat{a} \b{t} =\sqrt{\f{m \omega \b{t}}{2 \hbar}} \hat{Q}+\f{i}{\sqrt{2 m \omega \hbar} \b {t}} \hat{P}$ and $\hat{a}\b{t}^{\dagger}$. 
This deviation is a general consequence of non-adiabatic driving, independent of the model.
Such generation of coherence, associated with the bath, is a  unique property of the NAME.

In the Schr\"odinger frame the contribution to the coherence from the system-bath interaction are associated with the equations for the parameters $\beta\b t$ and $\gamma\b t$ (see below). 

\subsection{Instantaneous attractor }
The dynamics of the system, at each instant, can be imagined as motion toward a moving target, denoted as the {\em instantaneous attractor}.
The {\em instantaneous attractor} is defined as the local steady state, obtained by setting the LHS of Eq. \eqref{eq:betagamma} to zero.
This leads to:
\begin{equation}
\gamma_{ia}=0\,\,\,\,\,\,\text{and}\,\,\,\,\,\,\beta_{ia}=\text{log}\b{\f{k_{\uparrow}}{k_{\downarrow}}}=\text{log}\b{\f{N\b{\alpha\b t}}{N\b{\alpha\b t}+1}}~,
\label{eq:steadystate}
\end{equation}
and
\begin{equation}
\mean{\tilde{b}^{\dagger}\tilde{b}}_{ia} = N\b{\alpha \b t}~~~.
\label{eq:bdabbsteadystate}
\end{equation}
The {\em instantaneous attractor} is the temporal fixed point of the map and is an unattainable target as the system is continuously driven.

\begin{figure}[htb!]
\centering
\includegraphics[scale=0.27]{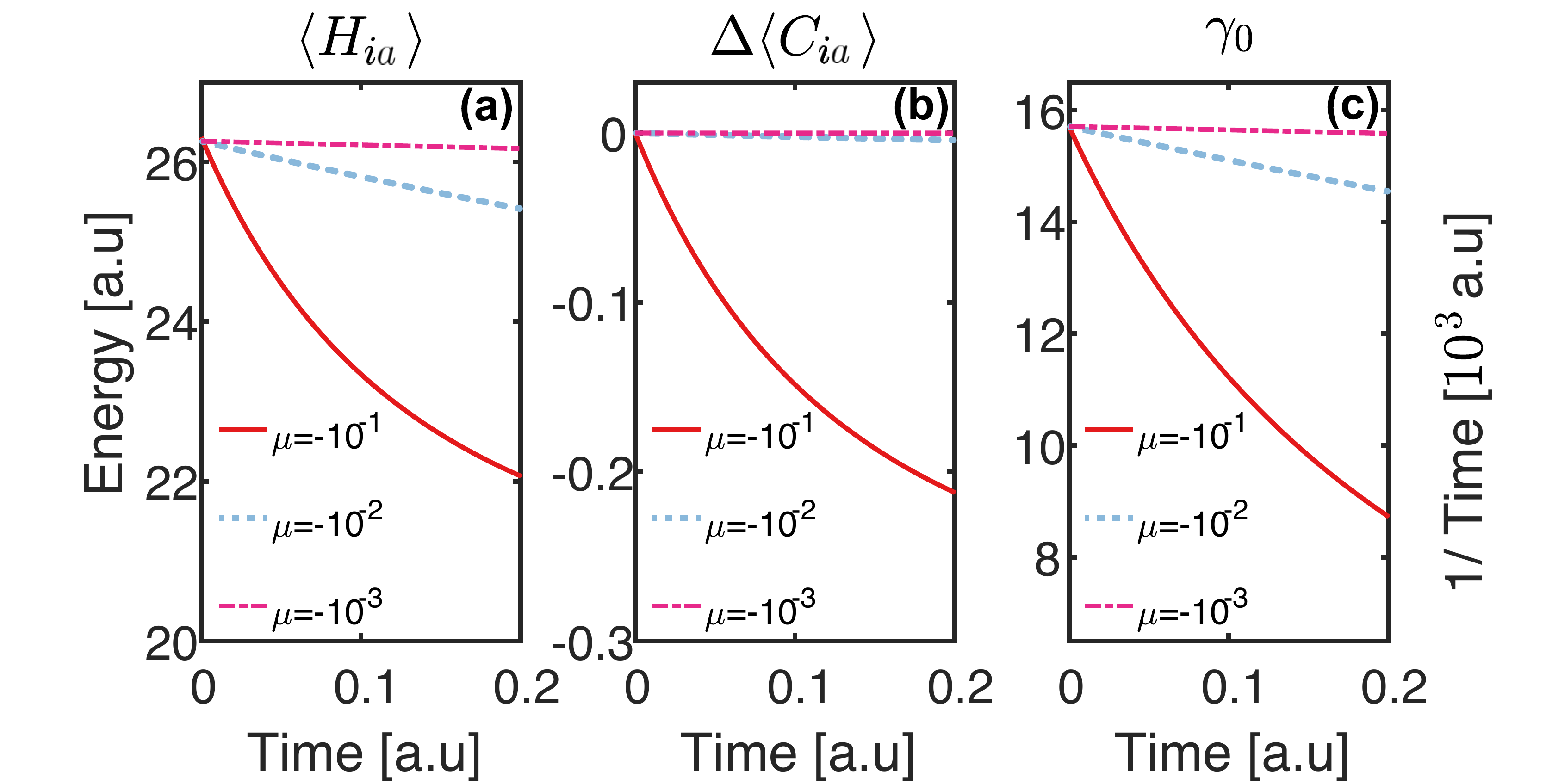} 
\caption{\label{HCss} (a) The energy value of the {\em instantaneous attractor} as a function of time,
for different values of the constant adiabatic parameter $\mu$. (b) The difference between the initial and the temporary value of the coherence of the {\em instantaneous attractor}, $\Delta\mean{\hat{C}_{ia}}=\mean{\hat{C}_{ia}\b {\tau}}-\mean{\hat{C}_{ia}\b {0}}$.  (c) The time-dependent rate coefficient (atomic units are used).
The initial frequency is $\omega \b 0 = 40$ and temperature of the bath $T = 20$, with initial values $\mean{\hat{H}\b 0}\approx \{ 26.3,26.2,26.2\} $ for $\mu=\{ 10^{-1},10^{-2},10^{-3}\}$ and similarly for $\mean{\hat{C}\b 0} = 1.31\cdot \{ 1,10^{-1},10^{-2}\}$.}
\end{figure}

The {\em instantaneous attractor} values for $\{\mean{\hat{H}_S},\mean{\hat{L}},\mean{\hat{C}},\mean{\hat{I}}\}$  are calculated by substituting Eq. (\ref{eq:steadystate}) in Eq. (\ref{eq:rho_S}) and utilizing  (\ref{eq:bdabbsteadystate}). 
We present the results for the {\em instantaneous attractor}, in Fig. \ref{HCss}, for different negative adiabatic parameters $\mu$. The harmonic oscillator's frequency decreases for $\mu<0$ leading to a decrease in the target energy $\mean{\hat{H}_{S}}_{ia}$. Coherence emerges via a non-vanishing  $\mean{\hat{C}}_{ia}$ arising from a finite driving speed (non-adiabatic). Figure \ref{HCss} shows that for vanishing $\mu$ the results coincide with the adiabatic solution, where the state follows the Hamiltonian and $\mean{\hat{C}}_{ia}\ra 0$. Similar generation of coherence has been obtained for a system coupled to a squeezed bath \cite{manzano2016entropy,li2016mutual}.
The {\em instantaneous  attractor} solution for  $\mean{\hat{L}}$ vanishes due to the independence of the steady state on $\gamma$. This result is independent of the parameter choice.

The dynamics can be viewed as motion in a time-dependent reference frame relative to a static bath.
In analogy to special relativity the bath observes a slowing down of the system frequency as $|\mu|$ is increased.
This modifies the rates which depend on the Fourier transform of the  correlations, with the system's frequency.
In addition, the non-adiabaticity of the system is equivalent to a system coupled to a squeezed bath. 
In the adiabatic  limit  ($\mu\ra0$) this effect vanishes and no coherence is generated.

\subsection{The asymptotic limit of the NAME}

The adiabatic limit is obtained when $\mu\ra0$. 
In this limit the operators $\hat{F}_{\pm}$, Eq. (\ref{eq:foperator}), converge to $\b{\hat{F}_{+},\hat{F}_{-}}\ra \sqrt{\f{\hbar}{2\omega\b 0m}}\b{\hat{a},\hat{a}^{\dagger}}$ while $\xi\b t\ra1$ and $\alpha\b t\ra\omega\b t$. 
Thus, in the adiabatic limit, Eq. (\ref{non_adi_HO}) reproduces the adiabatic Markovian Master equation as obtained by Albash et al. \cite{albash2012quantum},
\begin{equation}
\f d{dt}\hat{\rho}_{S}\b t= \b{\hat{\cal{U}}\b t+\gamma\b{\omega\b t}{\hat{\cal{D}}}\b t}\hat{\rho}_{S}\b t
\label{eq:adiabatic}
\end{equation}
where $\hat{{\cal{U}}}\b t \hat{\sigma}\equiv -\f{i}{\hbar}\sb{\hat{H}\b t,\hat {\sigma}}$ and 
\begin{multline}
   \hat{\cal{D}}\b t\hat{\sigma}\equiv  \hat{a}\b t\hat{\sigma}a^{\dagger}\b t-\f 12\{\hat{a}^{\dagger}\b t\hat{a}\b t,\hat{\sigma}\}\\+e^{-\hbar\omega\b t/k_B T}\b{\hat{a}^{\dagger}\b t\hat{\sigma}\b t \hat{a}-\f 12\{\hat{a}\b t\hat{a}^{\dagger}\b t,\hat{\sigma}\}} ~~. 
  \nonumber
\end{multline}
When $\omega$ is constant Eq. (\ref{eq:adiabatic}) becomes the standard Master equation of a thermalizing harmonic oscillator.

Comparing Eq. (\ref{eq:NAMEOH}) to the adiabatic Master equation (\ref{eq:adiabatic}) we notice two differences. First,  the decay rate is modified,
the non-adiabatic and adiabatic decay rates are related by
\begin{gather}
k_{\downarrow}=k_{\downarrow }^{adi}\f{J\b{\f{\kappa}2\omega\b t}\b{N\b{\f{\kappa}2\omega\b t}+1}}{J\b{\omega\b t}\b{N\b{\omega\b t}+1}}
\\
k_{\uparrow}=k_{\uparrow }^{adi}\f{J\b{\f{\kappa}2\omega\b t}N\b{\f{\kappa}2\omega\b t}}{J\b{\omega\b t}N\b{\omega\b t}}~~.
\nonumber
\end{gather}
 
For the case of Ohmic spectral density linear in the frequency as well as higher powers, $J(\omega) \propto \omega^n$
for $n \geq 1$, the non-adiabatic rate will be smaller than the adiabatic rate, due to $\f{\kappa}{2}\leq 1$.
It is important to note that the solution is valid when $|\mu|<2$ and  $\theta_{\pm} \in \mathbb{R}$. 
The point $|\mu|=2$ is an exceptional point representing the transition from damped to over-damped dynamics \cite{uzdin2013effects,moiseyev2013sudden}. 
Furthermore, $\mu$ and $\omega \b t$ are restricted by the secular approximation.

The NAME also differs in the jump operators $\hat{b},\hat{b}^{\dagger}$ vs. $\hat a,\hat a^{\dagger}$. 
In the adiabatic case:\\ $\hat a(t) =    \sqrt{\frac{m \omega(t) }{ 2\hbar }} \hat Q +i\frac{1}{\sqrt{ 2 m \hbar \omega(t)}} \hat P $, 
and in the non-adiabatic case
\begin{equation}
\hat b \b t=\sqrt{c}\b{A\hat{Q}\b 0+B\hat{P}\b 0}e^{i\theta_{+}\b t}
\label{eq:fexplicit}
\end{equation}
where $A$ and $B$ are defined below Eq. (\ref{eq:foperator}), $\sqrt{c}$ is the factor relating $\hat{b}$ and $\hat{F}_+$, and $\theta_+$ is given by Eq. \eqref{theta_pm}.
When $\mu \rightarrow 0$ Eq. (\ref{eq:fexplicit}) converges to the standard annihilation operator, $\hat{a}\b t$.

\section{Approximation analysis and regime of validity} 
\label{Approx validity sec}

We summarize the general derivation in section \ref{gen_deriv}, emphasizing the approximations performed and their range of validity. The relevant parameters of the composite system are the system-bath coupling strength $g$, the bath's spectral bandwidth $\Delta \nu$, the time-dependent quasi-Bohr frequencies $\{ \omega \b t \}$ of the system and the adiabatic parameter $\mu$, Eq. \eqref{mu_eq} \cite{mostafazadeh1997quantum}.

These four parameters determine four different timescales: 

\begin{enumerate}
\item{The system's typical timescale, $\tau_S=\text{max}_i\b{\frac{1 }{\omega_i \b t }}$, where $\omega_i$  are non-degenerate system Bohr frequencies.}
\item{ The timescale of the bath is defined by $\tau_B\sim \f{1}{\Delta \nu}$.} 
\item{ The relaxation time of the system, $\tau_R$, which is proportional to the coupling strength $\tau_R \propto g^{-2}$} \cite{albash2012quantum}.
\item{The timescale characterizing the rate of change of the system's energies due to the external driving, defined as $\tau_d$, the non-adiabatic timescale.}
\end{enumerate}

The microscopic derivation holds in the weak coupling limit, thus, terms of the order $O\b g^3$ and higher can be neglected (practically, only the even powers of $g$ will contribute, giving a correction of the order $O\b {g^4}$ to the derivation).
The Markov approximation is valid when  the  correlations decay rate is very fast relative 
to the coupling strength, leading to:
\begin{equation}
 g \tau_B \ll 1
\end{equation}
The next step is the secular approximation which  neglects the fast oscillating terms in Eq. (\ref{beforeRWA}). 
This approximation is valid for $\text{min}_t \sb{\theta_i \b {t} +\theta_j\b {t} }\gg 1$ when $\theta_i \neq - \theta_j$.

The non-adiabatic timescale $\tau_d$, is restricted by the timescale of the bath's correlations decay $\tau_B$. 
The timescale in which the driving field is changing should be slow relative to the bath's dynamics, i.e., $\tau_B\ll\tau_d$. In addition, the correlations decay fast relative to the system dynamics, $\tau_S \gg \tau_B$.
Here, $\tau_d$ can be evaluated by expanding $\theta_j \b {t-\tau_B}$ near the instantaneous time $t$ , (Cf. \ref{theta taylor exp}):
\begin{equation}
\theta_j\b{t-\tau_{B}}\approx\theta_{j}\b t-\theta'_{j}\b t\tau_{B}
\end{equation}
 Higher order powers can be neglected if, $|\theta_j^{\b{n+1}}\b t|\b{\tau_{B}}^{n+1}\ll|\theta_j^{\b n}\b t|\tau_{B}^{n}$, leading to $|\theta_j''\b t| \tau_B\ll|\theta_j'\b t|$. The typical timescale of the driving can be identified as
 \begin{equation}
 \tau_d \equiv \text{min}_{i,t}\sb {\theta_i'\b t/ \theta_i''\b t}~~. 
 \label{tau_d_eq}
 \end{equation}
A summery of the timescales is presented in Table \ref{table:timescales}.

\begin{table}
\centering{
\caption{Timescales}
\label{table:timescales}
\begin{tabular}{ |p{0.6cm}||p{3cm}|p{3cm}|  }
 \hline
 $\tau_S$     &  System intrinsic timescale & $\text{max}_{\omega_i}\b{1/{\omega_i \b t }}$\\
  $\tau_B$ & Bath correlation functions decay time& $\sim {1}/{\Delta \nu}$\\
    $\tau_R$ & Relaxation towards equilibrium lifetime& $\propto 1/{g^2 }$\\
    $\tau_d$ & Non-adiabatic timescale& $\text{min}_{i,t}\sb {\theta_i'\b t/ \theta_i''\b t}$\\
  \hline
\end{tabular}}
\end{table}

\subsection{Approximation analysis for the harmonic oscillator}

For the harmonic oscillator example $\tau_S\sim \f {1}{\omega\b t}$. In this case, the adiabatic parameter becomes $\mu =\f{\dot{\omega}}{\omega^2}$ and the non-adiabatic timescale is calculated with the help of Eq. \eqref{tau_d_eq} and \eqref{theta_pm} giving,  $\tau_d \sim {\omega \b t}/{\dot{\omega} \b t}=\b{\omega \b t \mu}^{-1}$. The adiabatic parameter and non-adiabatic timescale are related, however, the two differ in their physical implications. In contrast to the adiabatic approximation, which requires $\mu\ll1$,
the constraint on the non-adiabatic timescale, $\tau_d\gg\tau_B$ is dependent on the dynamics of the bath and allows for fast driving (large $\mu$), beyond the adiabatic regime.

The Born-Markov approximation conditions, $\tau_B \ll \tau_S$, $\tau_B \ll \tau_R$, leads to the following relations, $\omega \b t \ll \Delta \nu$ and $g\ll\Delta \nu$. 
Furthermore, the secular approximation leads to $\text{min}\,\omega\b t \gg \f{g^2}{\Delta \nu}$ and the 
driving protocol is restricted by $\mu \ll \text{min}\,\f{\Delta \nu}{\omega}$. 
Combining the inequalities above we can conclude  that the relevant system's frequency regime is
\begin{equation}
\f{g^2}{\Delta \nu}\ll \omega \b t \ll \Delta \nu\, \text{min}\sb{1,\mu^{-1}}~.
\label{omega_ineq}
\end{equation}

In the weak coupling limit for a bath with a constant and unbounded spectrum ($\Delta \nu \ra \infty$), the bath is delta correlated and the Master equation holds for any finite $\omega \b t$. 
Such a bath is hypothetical in practical scenarios, the bath's spectrum is finite and the validity regime defined by Eq. (\ref{omega_ineq}).

\section{Numerical analysis}
\label{Numerical_analysis_sec}
\begin{figure}[htb!]
\centering
\includegraphics[scale=0.27]{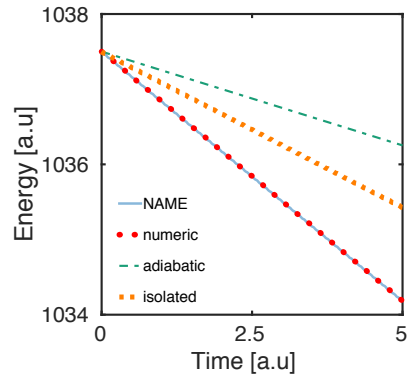} 
\caption{\label{comp plot}  The energy as a function of time for different solutions. The frequency decreases for a negative $\mu$ leading to a decrease in the energy. The initial state is of a Gibbs form: $\hat{\rho}_S=\exp \b{\beta \b{0} \hat{b}^{\dagger}\hat{b}}$. The model parameters are shown in a table in Appendix \ref{num:values}.  }
\end{figure}


We analyze the model by simulating numerically the system and bath.
The model is a driven harmonic oscillator coupled to 
a Bosonic bath. The bath consists of $N$ oscillators with an identical mass $m$ represented by the Hamiltonian
\begin{equation}
\hat H_B =\sum_{i=1}^{N}\bigg(\f{\hat p_{i}^{2}}{2m}+\f 12m\omega_{i}^{2}\hat q_{i}^{2}\bigg)~~.
\end{equation}
A linear system bath coupling is employed 
\begin{equation}
 \hat H_{I}=\omega \b t \hat Q\otimes\sum_{k=1}^N g_{k}\hat{q}_{k}  
\end{equation}
and a flat spectral density $J(\omega)=const$, in range $\omega\in \sb{\omega_{min},\omega_{max}}$. 
For the numerical compression we choose a different linear interaction than that in the analytical derivation, Eq. (\ref{eq:Intterm}), which simplifies the numerical calculations. The matching analytical derivation for the new interaction is modified accordingly.

The combined system, Eq. (\ref{sys_Ham}), and bath form a set of linear harmonic systems. Leading
to closed Heisenberg equations of motion for the set of operators 
$\hat P~,\hat Q,~ \hat P^2,~\hat Q^2,~\hat P \hat Q+\hat Q\hat P$ and for all $1\leq i\leq N$: $~\hat p_{i},~\hat q_{i}~,\hat p_{i}^2,~\hat q_{i}^2 ,~\hat p_{i} \hat q_{i}+\hat q_{i} \hat p_{i}$.
We solve for the expectation values of the operators  and the solution for the system's variables is  translated to the set of operators $\mean{\hat H_S \b t}$, $\mean{\hat L\b t}$ and $\mean{\hat C\b t}$.

In the limit when the number of the bath modes diverge, $N\ra\infty$,   $\omega_{max} \rightarrow  \infty$, the numerical approximation converges to the NAME's solution.
The equations of motion were solved for the second moments by a Dormand-Prince Runge Kutta method (DP-RK4) with a constant time step, see  Appendix \ref{appendix:num} for more details.

In Fig. \ref{comp plot}, the energy as a function of time is compared for the  adiabatic, isolated, NAME and numerical solutions. 
The results show a good match between the NAME and the independent numerical approach, 
while the adiabatic and isolated solutions deviate substantially from the expected energy change. Hence, the numerical result verifies the analytical derivation and solution for the NAME.  
To see this effect in the numerical simulation $\mu \omega$ should be comparable to the decay rate.    
In contrast, when $\mu$ is large the free propagation dominates.

\section{Conclusion}
\label{Conclusion}

Then Non-Adiabatic Master Equation (NAME)  addresses the issue of the environment's effect on the dynamics of a driven quantum system. This Master equation generates a Markovian reduced description for a driven quantum system going beyond the adiabatic framework.  
The equation is cast into the form of a time-dependent Gorini-Kossakowski-Lindblad-Sudarshan equation (GKLS) where both the operators and the kinetic coefficient are time-dependent.

A condition necessary to derive the NAME is a Lie algebra of operators which span both the driven and bare Hamiltonian and the system-bath coupling operators. This allows to obtain the free propagator and the time-dependent jump operators. These are identified as the eigenoperators of the propagator, Eq. \eqref{F_def_eq}. Furthermore, for the equation to be valid we require a timescale separation between the system and driving timescales, and the bath's correlation time.

The  NAME incorporates as limits, the time-independent, periodically driven  and the adiabatic Master equations. In comparison with the adiabatic \cite{albash2012quantum} or post adiabatic \cite{yamaguchi2017markovian} Master equations, the NAME mixes population
and coherence. The differences can be traced to the form of the jump operators, Eq. (\ref{non_adiabatic_master_eq}), 
composing the time-dependent GKLS equation. In the adiabatic case the jump operators are eigenoperators of the instantaneous Hamiltonian, in contrast, in the NAME the jump operators are eigenoperators of the free propagator.

Using the NAME we are able to explicitly solve the problem of a time-dependent harmonic oscillator coupled to a bath, Sec. \ref{HO_expample_sec}. 
The solution is facilitated by choosing a driving protocol dictated by a constant adiabatic parameter $\mu$. The SU(1,1) Lie algebra is employed to derive the Master equation and to represent the system as a generalized Gibbs state in the operators of the algebra.  This form is equivalent to a squeezed thermal state and enables the explicit solution of the dynamics.
Such restriction of a constant $\mu$ can be uplifted by using a piecewise approach, 
decomposing an arbitrary protocol to small time intervals with a constant $\mu$. 
 
 The driven harmonic oscillator model exhibts reduced decay rates in the NAME compared to the rates obtained from the adiabatic Master equation. The reason is an effective reduction of the system frequency $\alpha \b t<\omega\b t$ as seen by the bath. The explicit solution demonstrates the mixing of coherence and population in the equations of motion. Furthermore, when solving the dynamics of the NAME in the Schr\"odinger picture, the {\em instantaneous attractor} can be identified. At each instant, the dynamics directs the system towards the {\em instantaneous attractor}. Coherence is generated since the {\em instantaneous attractor} is not diagonal in the instantaneous energy basis.

The dynamics of the NAME is compared to a numerical simulation.  
The simulation converges to the analytical prediction of the NAME.

The NAME addresses the problem of a driven open system within the Markovian approximation. In any control problem of open quantum systems this is the typical scenario, \cite{koch2016controlling,levy2018noise,vinjanampathy2018speeding}. Such a control problem appears abundantly in annealing approaches to quantum computing \cite{das2008colloquium} and in quantum gates \cite{jaksch2000fast,duan2001geometric,jonathan2000fast,nielsen2002quantum}.

\section{Acknowledgement}
 We thank KITP for the hospitality. This research was supported in part by the National Science Foundation under Grant No. NSF PHY-1748958. We thank Robert Alicki and Luis A. Correa for fruitful discussions. We also acknowledge the support of the Israel Science Foundation, number 2244/14.
\appendix

\section{Eigenoperators}
\label{App:eigenoperators}
We assume the system dynamics can be described by a time-independent operator basis   $\{\hat{G}\}$ including a finite number of operators which are elements of a Lie algebra
\begin{equation}
\sb{\hat{G_j},\hat{G_i}} = \sum_{k=1}^N c_{k}^{ij} \hat{G}_k~~,
\label{eq:B sum}
\end{equation}
where $c_k^{ij}$ are the structure constants.  

If the Hamiltonian $\hat{H}_S\b t$ at initial time is a linear combination of the operators $\{\hat{G}\}$, it is a member of the algebra and can be expressed as:
\begin{equation}
\hat{H_S}\b t = \sum_{j=1}^N h_j \b t \hat{G}_j~~.
\label{eq:H sum}
\end{equation}
With the help of the identity Eq. \eqref{eq:H sum} and the Heisenberg Equation one concludes that the equations of motion for the system operators are closed under the Lie algebra. In addition, for any closed Lie algebra the time evolution operator can be written as \cite{alhassid1978connection}:
\begin{equation}
 \hat{U}_S \b t = \displaystyle\prod_j^N e^{r_j\b t \hat{G}_j}~~,
 \label{apeq:Us}
 \end{equation}
where $r_j \b t$ are time-dependent coefficients.

The eigenoperators can be found by representing the dynamics in Liouville space (known also as Hilbert-Schmidt space). Such Hilbert space is a state space of system operators, $\{\hat{X} \}$, endowed with an inner product defined by, $\b{\hat{X_i},\hat{X_j}}\equiv\text{tr}\b{\hat{X_i}^{\dagger}\hat{X_j}}$ \cite{am2015three,von2018mathematical,horn1990matrix}.

In the Liouville representation, the system's dynamics are calculated in terms of a chosen basis of operators spanning the Liouville space, (such as $\{ \hat{G}\}$). 
This basis of operators construct a vector $\v{v} \b t$ in observable space. For example, the dynamics of a two-level-system is described in the Bloch sphere where the basis is constructed from the Pauli operators.

Employing the Heisenberg equation of motion, the dynamics of $\v v$ is given by, 
 \begin{equation}
 \f{d}{dt}\v{v} \b t = \b{\f{i}{\hbar} \sb{\hat{H}\b t,\cdot} +\pd{}{t}} \v{v}\b t 
 \label{dynamics Liouv}
 \end{equation}
Here, we consider a finite basis of size $N$, which also forms a closed Lie algebra. This guarantees that the Heisenberg equations of motion \eqref{dynamics Liouv} can be solved within the basis \cite{alhassid1978connection}, implying that Eq. \eqref{dynamics Liouv} can be represented in a vector matrix form,
\begin{equation}
    \f{d}{dt}\v{v} \b t = {\cal{G}} \b t \v v \b t~~,
    \label{appendix:heis}
\end{equation}
where  ${\cal{G}}\b t$ is an $N$ by $N$ matrix and $\v{v}$ is a $N$-dimensional vector.
For a hermitian Hamiltonian the algebraic properties insure that  ${\cal{G}}\b t$ is diagonalizable, see following subsection. Let $\{\v{F}\b t \}$ be the eigenvector basis of ${\cal{G}}\b t$ and $\v{F}_j \b t\in\{\v{F}\b t \}$, then $\v{F}_j \b t$ satisfies the instantaneous eigenvalue equation
\begin{equation}
    {\cal{G}} \b t \v{F}_j \b t= \chi_j\b t \v{F}_j \b t~~,
    \label{app:eig}
\end{equation}
where $\chi_j \b t\in \mathbb{C}$.
Using Eq. \eqref{app:eig} into Eq. \eqref{appendix:heis} and solving for  $\v{F}_j \b t$ leads to,
\begin{equation}
 \v{F}_j \b t ={\cal{U}}_S\b t \v{F}_j \b 0 = e^{\chi_j \b t}\v F_j \b 0~~,
 \label{app:F_jeig}
\end{equation}
where ${\cal{U}}_S\b t$ is the propagator in Liouville space. We define $\lam_j \b t\equiv e^{\chi_j \b t}$  and represent Eq. \eqref{app:F_jeig} in the wave-function Hilbert space to obtain:
\begin{equation}
     \tilde{F}_j\b t=\hat{U}_S^\dagger \b t \hat{F}_{j}\b 0  \hat{U}_S \b t=\lam_j \b t \hat{F}_{j}\b 0 ~~.
\end{equation}
This is the eigenvalue equation for the eigenoperators (identical to Eq. \eqref{F_def_eq}).

\subsection{Diagonalazability of ${\cal{G}}\b t$}
\label{sec:diagonal}
The dynamics of the density matrix is given by the Liouville von-Neumann equation, $\f{d\hat{\rho}_S\b t}{dt}=-\f{i}{\hbar}\sb{\hat{H}_S\b t,\hat{\rho}_S\b t}$.
By preforming a vec-ing procedure, the density matrix $\hat{\rho}_S$ ($N$ by $N$ matrix) is represents as a $N^2$ vector, $\v{r}$ \cite{am2015three,machnes2014surprising,horn1990matrix}. This is equivalent to choosing the representation basis in Liouville space as the set of matrices with all-zero entries, except one. Following the derivation presented in \cite{am2015three}, the Liouville von-Neumann super-operator can be represented as a $N^2 \times N^2$ matrix in Liouville space,
\begin{multline}
    -\f{i}{\hbar}\sb{\hat{H}_S\b t,\hat{\rho}_S\b t}\ra\\ -\f{i}{\hbar}\b{I\otimes \hat{H}_S\b t-\hat{H}_S\b t^T \otimes I}\v r \equiv {\cal{D}}\b t\v r
\end{multline}
where $\otimes$ is the Kronecker direct product.  The Hamiltonian in the diagonal form, implying that the matrix ${\cal{D}}\b t$ is diagonalizable. Transforming to a different basis (can be time-dependent) in Liouville space involves transformation matrices. The transformation to a time-dependent basis is only a change in the representation, therefore we assume  that also ${\cal{G}}\b t$ is diagonalizable. Thus, the diagonalability property is invariant to the change of basis.

As a product of diagonalizable matrices leads to a diagonalizable matrix, the generator ${\cal{G}}\b t$ in Eq. \eqref{appendix:heis} is diagonalizable.

\section{Expanding the interaction operator $\tilde{A}_k$ using the Lie algebra structure}
\label{sec:expansion set}

The jump operators are eigenoperators of the free evolution obeying, Eq. \eqref{F_def_eq}. They form a complete basis within the system's algebra.
Equation \eqref{apeq:Us} implies that operators in the interaction representation are also closed under the free propagation.

  If the operator $\hat{A}_k$ (Eq. \eqref{eq:H_I}) is also an element of the Lie algebra, it can be expanded in the interaction representation in terms of the set $\{\tilde{F}_j\b t\}$,
 \begin{equation}
 \tilde{A}_k \b t = \sum_j \chi_j^k \b t \tilde{F}_j\b t~~~.
 \label{eq:Adecomp}
 \end{equation}
 The coefficients $\chi_j^k \b t $ are in general complex, therefore, can be written in a polar representation leading to the desired form: $ {\tilde{A}}_{k}\b t=\sum_{j}\xi_{j}^{k} \b t e^{i\theta_{j}^k\b t} {\hat{F}}_{j}\b 0$ (Eq. \eqref{Decomposition to eigen operators}). Here, $\xi_{j}^{k} \b t = |\chi_j^k \b t\cdot\lam_j \b t |$  and $\theta_j^k \b t=\phi_j \b t+\text{arg}\b{\lam_j \b t}$.
 The amplitude $\xi_j^k \b t$ of a complex number is necessarily positive, leading to positive decay rates in the NAME (\ref{non_adiabatic_master_eq}).



\section{Generalized Gibbs state}
\label{app:GGS}

In section \ref{subsec:Solution_NAME} the NAME is derived for the open system dynamics of a parametric harmonic oscillator 
employing  a solution that at all times can be described as a squeezed Gaussian state (ensemble) \cite{jaynes1957information,langen2015experimental}. 
This solution is a special case obtained when the system can be described in terms of a Lie algebra of operators. In such a case,  
the state of the system at all times is represented as a generalized Gibbs state (GGS). 
The GGS is determined by maximum entropy with respect to the set of observables $\{\langle \hat{X} \rangle\}$ where
the operators $\hat X$ are members of the Lie algebra .The state has the form:
\begin{equation}
 \hat{\rho}_S\b t = \displaystyle e^{ \sum_j \lambda_j\b t \hat{X}_j}~~,
\label{eq:rho_maxentrop}
\end{equation} 
where $\lambda_j$ are Lagrange multipliers.

To maintain this form, the set of operators $\{\hat{X}\}$ has to be  closed under the dynamics generated by the equation of motion. 
Using the Lie algebra properties, the state can written in a product form in terms of  the set $\{\hat{X}\}$, \cite{kosloff2017quantum,alhassid1978connection,wei1964global},
\begin{equation}
 \hat{\rho}_S\b t = \displaystyle\prod_i^N e^{d_j\b t \hat{X}_j}~~,
\label{eq:rho_decomp2}
\end{equation} 
where $d_j \b t$ are time-dependent coefficients.

The squeezed Gaussian state, assumed in Sec. \ref{subsec:Solution_NAME} is a special case of a generalized Gibbs state. 
Accordingly, a solution of the dynamics follows the derivation in \ref{subsec:Solution_NAME} 
obtaining a set of coupled differential equations similar to Eq. (\ref{eq:betagamma}) 
which can be solved by analytical or numerical methods. 

\section{Derivation of the Master equation up to first order in the bath's correlations decay time}
\label{sec:higher_order}

In section \ref{gen_deriv}  the NAME, Eq. (\ref{non_adiabatic_master_eq}) is derived, assuming the bath's correlation decay timescale is shorter than the system and driving timescale. The derivation involves the lowest possible order which captures the effect of the non-adiabatic driving and  is exact for a delta-correlated bath. However, 
in realistic scenarios the bath is characterized by a finite spectral width and therefore has a non-vanishing bath correlation time, $\tau_B$, which defines the range of validity. It is possible to go beyond the lowest order correction introduced in Eq. (\ref{non_adiabatic_master_eq}), and include higher order corrections in $\tau_B$. In the following section we present a derivation of the NAME for the harmonic oscillator including the first higher order correction, an extension for the general case is straightforward.

The starting point of the derivation of the NAME is the Markovian quantum master equation, \citep{breuer2002theory}, (Eq. (3.118) p.132):
\begin{multline}
\f d{dt}\tilde{\rho}_{S}\b t=\\-\f{1}{\hbar^2}\int_0^{\infty}{ds\,\text{tr}_{B}\sb {\tilde{H}\b t,\sb{\tilde{H}\b{t-s},\tilde{\rho}_{S}\b t\otimes\hat{\rho}_{B}}}}~~~.
\label{eq:Markovian_master}
\end{multline}
The Hamiltonian in the interaction picture is first decomposed in terms of the set of eigenoperator:
\begin{multline}
\tilde{H}\b t=i \xi\b t\sum_{j=\pm}\hat{F}_{j}e^{i\theta_{j}\b t}\sum_{k}g_{k}\sqrt{\f{\hbar m\omega_{k}}{2}}\\ \times \b{\hat{b}_{k}^{\dagger}e^{i\omega_{k}t}-\hat{b}_{k}e^{-i\omega_{k}t}}~,
\label{eq:int_H}
\end{multline}
where $\hat{F}_j \equiv \hat{F}_j \b 0$.
Equation (\ref{eq:Markovian_master}) includes terms of the form $\text{tr}_{B}\sb{\tilde{H}\b t\tilde{H}\b{t-s}\tilde{\rho}_{S}\b t\otimes\hat{\rho}_{B}}$.
Next, we demonstrate how such a term is calculated explicitly using Eq. (\ref{eq:Markovian_master}). Contribution of other terms to the Master equation can be calculated in a similar manner. 

\begin{multline}
\text{tr}_{B}\sb{\tilde{H}\b{t-s}\b{\tilde{\rho}_{S}\b t\otimes\hat{\rho}_{B}}\tilde{H}\b{t}}=\\
-\xi\b t\xi\b{t-s}\f{\hbar m}{2}\sum_{i,j}\sum_{k,k'}\sqrt{{\omega_{k} {\omega_{k'}}}}g_{k}g_{k'}\hat{F}_{i}\tilde{\rho}_{S}\b t\hat{F}_{j}\\\times
e^{i\theta_{i}\b{t-s}}e^{i\theta_{j}\b{t}}
\sum_{k}\text{tr}_{B}\big[\b{\hat{b}_{k}^{\dagger}e^{i\omega_{k}t}-\hat{b}_{k}e^{-i\omega_{k}t}}\\\times
\b{\hat{b}_{k'}^{\dagger}e^{i\omega_{k'}\b{t-s}}-\hat{b}_{k'}e^{-i\omega_{k'}\b{t-s}}}\hat{\rho}_{B} \big]~~.
\label{eq:xsi_long}
\end{multline}
We proceed by expanding $\theta_{i}\b{t-s}$ near $s=0$.  In the range of validity determined by the decay of the correlation $s \sim \tau_B$ or $s<\tau_B$, allowing to approximate   $s^{2}\approx\tau_{B}s~$, then: 
\begin{equation}
\theta_{i}\b{t-s}\approx\theta_{i}\b t-\theta_{i}'\b ts+\f{\theta_{i}''\b t}2\tau_{B}s~~.
\label{eq:theta_exp2}
\end{equation}
We define $\bar{\alpha}\b t\equiv-\theta_{i}'\b t+\f{\theta_{i}''\b t}2\tau_{B}$. The definition of $\bar{\alpha}\b t$ is similar to the definition in Eq. (\ref{theta taylor exp}) for the first order expansion in $s$.

Substituting Eq. (\ref{eq:theta_exp2}) into Eq. (\ref{eq:xsi_long}) and performing the secular approximation terminates terms for which $\theta_{i}\b t\neq-\theta_{j}\b t$. In addition, for a Bosonic bath in thermal equilibrium $\mean{\hat b_{k}\hat b_{k}}=\mean{\hat b_{k}^{\dagger}\hat b_{k}^{\dagger}}=0$, $\mean{\hat b_{k}\hat b_{k'}}=\delta_{k,k'}$, and Eq. (\ref{eq:xsi_long}) is simplified to the form 
\begin{multline}
\text{tr}_{B}\sb{\tilde{H}\b{t-s}\b{\tilde{\rho}_{S}\b t\otimes\hat{\rho}_{B}}\tilde{H}\b{t}}\\
=\f{\hbar m}{2}\xi\b t\xi\b{t-s}\sum_{i=\pm}\hat{F}_{i}  \tilde{\rho}_{S}\b t\hat{F}_{i}^{\dagger}e^{i\bar{\alpha}_{i}\b ts}\times\\
\sum_{k}{\omega_{k}}g_{k}^{2}\b{\mean{\hat{b}_{k}^{\dagger}\hat{b}_{k}}
e^{i\omega_{k}s}+\mean{\hat{b}_{k}\hat{b}_{k}^{\dagger}}e^{-i\omega_{k}s}}~.
\end{multline}
The coefficients $g_k$ have units of inverse time. Thus, in the continuum limit, the sum over $g_k^2$ can be replaced by an integral:
\begin{equation}
\sum_{k}g_{k}^{2}\ra\int_0^{\infty}{f\b{\omega_k} \chi \b{\omega_k} d\omega_k}~~,
\label{eq:continua}
\end{equation}
where $f\b{\omega}$ is the density of states, such that $f\b{\omega}d\omega$ gives the number of oscillators with frequencies in the interval $\sb{\omega,\omega + d\omega}$ \cite{carmichael2009open},  and $\chi \b{\omega}$ is the coupling strength function. On the LHS of Eq. (\ref{eq:continua}) the variable $k$ is an integer while on the RHS it designates the wave number which is a continuous variable in the continuum limit. The two functions define the spectral density function $J \b{\omega} = f\b{\omega} \chi \b{\omega}$, leading to:
\begin{multline}
\text{tr}_{B}\sb{\tilde{H}\b{t-s}\b{\tilde{\rho}_{S}\b t\otimes\hat{\rho}_{B}}\tilde{H}\b{t}}=\\
\xi\b t\xi\b{t-s}\sum_{i=\pm}\hat{F}_{i}\tilde{\rho}_{S}\b t\hat{F}_{i}^{\dagger}e^{i\bar{\alpha}_{i}\b ts}\times\\
\int_0^{\infty}{d\omega_{k}}{\omega_{k}}J\b{\omega_{k}}\f {\hbar m}{2 }\b{\mean{\hat{b}_{k}^{\dagger}\hat{b}_{k}}e^{i\omega_{k}s}+
\mean{\hat{b}_{k}\hat{b}_{k}^{\dagger}}e^{-i\omega_{k}s}}\\+\text{similar terms}~.
\label{eq:derivation1}
\end{multline}
By inserting Eq. (\ref{eq:derivation1}) in the Markovian quantum master equation we obtain the reduced dynamics  
\begin{multline}
\f d{dt}\tilde{\rho}_{S}\b t= \sum_{i=\pm}\hat{F}_{i} \tilde{\rho}_{S}\b t
\hat{F}_{i}^{\dagger}\int_0^{\infty}{d\omega_{k}}{\omega_{k}}J\b{\omega_{k}}\f {m}{2\hbar}\xi\b t \times \\
\int_0^{\infty}{ds\xi\b{t-s}e^{i\bar{\alpha}_{i}\b ts}\b{\mean{\hat{b}_{k}^{\dagger}\hat{b}_{k}}e^{i\omega_{k}s}
+\mean{\hat{b}_{k}\hat{b}_{k}^{\dagger}}e^{-i\omega_{k}s}}}\\+\text{similar terms}~.
\label{eq:additional}
\end{multline}
Assuming the change in $\xi$ is slow relative to the decay of the bath correlation functions then 
\begin{equation}
\xi\b{t-s}\approx\xi\b t-\xi'\b t\tau_{B}~~.
\end{equation}
We define
\begin{multline}
  \bar{\Gamma}\b t\equiv\f m{2 \hbar }\int_0^{\infty}{d\omega_{k}}{\omega_{k}}J\b{\omega_{k}}\times\\\int_0^{\infty}{dse^{i\bar{\alpha}_{i}\b ts}\b{\mean{\hat{b}_{k}^{\dagger}\hat{b}_{k}}e^{i\omega_{k}s}
+\mean{\hat{b}_{k}\hat{b}_{k}^{\dagger}}e^{-i\omega_{k}s}}}~~.  
\end{multline}
Decomposing $\bar{\Gamma}\b t$ to a real and pure imaginary part and using the identity $\int_0^{\infty}{ds\,e^{-i\eps s}}=\pi\delta\b{\eps}-i \mathcal{P}
\frac{1}{\eps}$ (here $\delta\b x$ is the Dirac delta function and $\mathcal{P}$ is the Cauchy principle value we obtain
\begin{equation}
\bar{\Gamma}\b t\equiv\b{\f 12\gamma\b{\bar{\alpha}_{i}\b t}+iS\b{\bar{\alpha}_{i}\b t}}~~,
\end{equation}
where
\begin{equation}
\gamma\b{\bar{\alpha}_{i}\b t}=\f{m \pi}{ \hbar}{\bar{\alpha}_{i}\b t}J\b{\bar{\alpha}_{i}\b t}\b{\bar{N}\b{\bar{\alpha}_{i}\b t}+1}~~,
\end{equation}
and
\begin{equation}
S\b{\bar{\alpha}_{i}\b t}={\cal{P}}\sb{\int_0^{\infty}{d\omega_{k}\sb{\f{1+\bar N\b{\omega_{k}}}{\bar{\alpha}_{i}\b t-\omega_{k}}+\f{\bar N\b{\omega_{k}}}{\omega_{k}+\bar{\alpha}_{i}\b t}}}}~~. 
\end{equation}

An identical derivation is carried out for the additional terms in  Eq. (\ref{eq:additional}). After some algebra the first order correction to  the NAME  is obtained:
\begin{multline}
\f d{dt}\tilde{\rho}_{S}\b t=\b{|\xi\b t|^{2}-\xi\b t\xi'\b t\tau_{B}}\\\sum_{i}\gamma\b{\bar{\alpha}_{i}\b t}\b{\hat{F}_{i}\tilde{\rho}_{S}\b t\hat{F}_{i}^{\dagger}-\f 12\{\hat{F}_{i}^{\dagger}\hat{F}_{i},\tilde{\rho}_{S}\b t\}}~~.
\end{multline}

For the harmonic oscillator example, the derivatives of $\theta_i$ can be calculated from Eq. (\ref{theta_pm}) 
$\theta_{\pm}'\b t=\mp\f{\kappa\omega\b t}2$ and $\theta_{\pm}''\b t=\mp\f{\kappa\mu\omega^{2}\b t}2$.
Leading to:
\begin{equation}
\bar{\alpha}_{+}\b t\equiv \f{\kappa\omega\b t}2\b{1-\f{\mu\omega\b t}2\tau_{B}}~~. 
\label{ap:alpha bar}
\end{equation}
Notice here that this expression is the first order correction to $\alpha_{+}\b t$, (introduced for the general case  in equation (\ref{theta taylor exp}) and is derived for the harmonic oscillator from $\theta_+$, Eq. \eqref{theta_pm}).\\
The harmonic oscillator NAME, in the interaction representation, including the first order correction is of the form
\begin{multline}
\f d{dt}\tilde{\rho}_S\b t=\b{|\xi\b t|^2+\mu{\tau_{B}\omega\b 0}/2}\gamma\b{\alpha_{+}
\b t}\times \\
\Bigg(\hat{F}_{+}\tilde{\rho}_{S}\hat{F}_{-}-\f 12\{\hat{F}_{-}\hat{F}_{+},\tilde{\rho}_{S}\b t\}\\+e^{-\bar{\alpha}_{+}\b t/k_B T}\b{\hat{F}_{-}\tilde{\rho}_{S}\hat{F}_{+}-\f 12\{\hat{F}_{+}\hat{F}_{-},\tilde{\rho}_{S}\}}\Bigg)~~~.
\label{eq:high}
\end{multline}

Two differences appear between Eq. (\ref{eq:high}) and the lower order derivation: First, there is a small correction to the decay rate in the order of $\mu \tau_B \sim \omega\b t \f{\tau_B}{\tau_d}$ (where $\tau_B\ll \tau_d$). The negligible correction justifies the approximation performed in Eq. \eqref{beforeRWA}, in the main derivation, where only the zeroth order of $\xi \b{t-s}$ in $s$ is kept.  Second, a memory-like correction arises due to the phase higher order correction. The  higher order term in $\bar{\alpha}_+$, Eq. \eqref{ap:alpha bar} is proportional to $-\mu$, and therefore decreases or increases $\bar{\alpha}_+$ depending on the sign of $\mu$. For spectral density $J\propto \omega ^r$ where $r\geq1$ this will lead to a decay rate which is retarded in time. The effect can be understood as a delay in the reaction of the bath to the system's change in time.
This effect will increase when the correlation time increases and vanishes for a delta correlated bath.

\section{Free propagation}
\label{Adiabatic propagation}
The unitary dynamics of the operators $\{\hat{H}_S\b t$, $\hat{L}\b t,\hat{C}\b t\}$ are given by \cite{kosloff2017quantum}:
\begin{equation}
\label{eq:Adiprop}
{{\cal U}_S}\b t=\f 1{\kappa^2}\f{\omega\b t}{\omega\b 0}\left[{\begin{array}{cccc}
4-\mu^{2}c & -\mu\kappa s & -2\mu\b{c-1} & 0\\
-\mu\kappa s & \kappa^{2}c & -2\kappa s & 0\\
2\mu\b{c-1} & 2\kappa s & 4c-\mu^{2} & 0\\
0 & 0 & 0 & 1
\end{array}}\right]
\end{equation}
where $\kappa = \sqrt{4-\mu^2}$ and $c=cos\b{\kappa \theta \b t}$, $s=sin\b{\kappa \theta \b t}$.
The free propagation mixes coherence and populations due to driving \cite{kosloff2017quantum}.

\section{Expanding the interaction term for the harmonic oscillator model in terms of eigenoperators of the propagator}
\label{Appendix:exp}
We define two new time-dependent operators, $\bar{Q}\b t=\sqrt{\omega\b t}\hat{Q}$ and $\bar{P}\b t=\f 1{m\sqrt{\omega\b t}}\hat{P}$, leading to equations of motion which can be written in a matrix vector notation,
\begin{equation}
\f d{dt}\sb{\begin{array}{c}
\bar{Q}\\
\bar{P}
\end{array}}=\omega\b t\sb{\begin{array}{cc}
\f{\mu}2 & 1\\
-1 & -\f{\mu}2
\end{array}}\sb{\begin{array}{c}
\bar{Q}\\
\bar{P}
\end{array}}~~.
\end{equation}
Diagonalizing the constant matrix leads to eigenoperators which are associated with the left eigenvectors of the constant matrix,
\begin{equation}
\hat{u}_{\pm}=\f 12\b{\mu\pm i\kappa}\bar{Q}+\bar{P},
\label{eq:u}
\end{equation}
which propagate in time as  $\hat{u}_\pm \b t = \hat{u}_\pm \b 0 e^{i \theta_{\pm}}$.
Here, $\theta_{\pm}\equiv\pm \f{\kappa}{2}\int_0^t dt' \omega \b t'$.

By defining $\hat{F}_{+}\equiv \f{i}{\kappa \omega \b 0}\hat{u}_{-}\b 0$, and utilizing equations \eqref{eq:u} and the definition of $\bar{Q}\b t$, we obtain the decomposition,

\begin{equation}
\hat{Q}\b t=\sqrt{1-\mu\omega\b 0t} \b{\hat{F}_{-}e^{i\theta_{-}}+\hat{F}_{+}e^{i\theta_{+}}} ~~.
\end{equation}

Defining $\xi \b t\equiv \sqrt{1-\mu\omega\b 0t} $ leads to the desired form.


\section{Calculation of the expectation values for $\hat{H}_S\b t$, $\hat{L}\b t$, $\hat{C}\b t$} 
\label{sec:HLCexp}
We define a vector in Liouville space $\v v\b t=\{\hat{H}_S\b t,\hat{L}\b t,\hat{C}\b t,\hat{I}\}^{T}$ similarly to the derivation in \cite{kosloff2017quantum}.

The dynamics of the isolated system is given by:
\begin{equation}
\v v\b t={{\cal U}_S}\b{t}\v v\b 0
\label{eq:vUb0}
\end{equation}
where ${\cal{U}}_S \b t$ is given in Eq. (\ref{eq:Adiprop}).

The operators of $\v v\b 0$ can be written in terms of the basis $\v b\b 0=\{\hat{b}^{2}\b 0,\hat{b}^{\dagger}\hat{b}\b 0,\hat{b}^{\dagger2}\b 0\}^{T}$, the transformation is summarized in the matrix form by
\begin{equation}
\v v\b 0={\cal M}\v b\b 0 ~~.
\label{eq:MatixM}
\end{equation}
where ${\cal{M}}={\cal{M}}_1\b{{\cal{M}}_2}^{-1}$, ${\cal{M}}_{1,2}$ are given by:
\begin{equation}
{\cal{M}}_1=\sb{\begin{array}{cccc}
\f 12m\omega^{2} & \f 1{2m} & 0 & 0\\
-\f 12m\omega^{2} & \f 1{2m} & 0 & 0\\
0 & 0 & \f{\omega}2 & 0\\
0 & 0 & 0 & 1
\end{array}}
\end{equation}
\begin{equation}
    {\cal{M}}_2=\tilde{c}\sb{\begin{array}{cccc}
A^{2} & B^{2} & 2Re\b{A^{*}B} & 0\\
|A|^{2} & |B|^{2} & 2Re\b{A^{*}B} & 2Im\b{A^{*}B}i\hbar\\
A^{*2} & B^{*2} & A^{*}B^{*} & 0\\
0 & 0 & 0 & 1
\end{array}}
\nonumber
\end{equation}
with $\tilde{c} = \b{2\hbar \text{Im}\b{A^* B}}^{-1}$.

Inserting Eq. (\ref{eq:MatixM}) into Eq. (\ref{eq:vUb0}) and defining ${\cal T}\equiv{\cal U}_S{\cal M}$ we obtain 
\begin{equation}
\v v\b t={\cal T}\v b\b 0
\end{equation}
and for the expectation values
\begin{equation}
\mean{\v v\b t}={\cal T}\mean{\v b\b 0}~~~.
\end{equation}
The expectation values of the basis $\v b\b 0$ are calculated using Eq. (\ref{eq:rho_S}):

\begin{equation}
\mean{\tilde{b}^{\dagger}\tilde{b}}=  \text{tr}\b{\tilde{b}^\dagger \tilde{b} \b 0  \tilde{\rho}_S\b t}=\f{\b{e^{-2\beta}-4|\gamma|^{2}-1}}{2\b{\b{e^{-\beta}-1}^{2}-4|\gamma|^{2}}}-\f 12
\end{equation}
and
\begin{equation}
\mean{\tilde{b}^{2}}=\b{\mean{\tilde{b}^{\dagger 2}}}^*=\f{2\gamma^{*}}{\b{e^{-\beta}-1}^{2}-4|\gamma|^{2}}~~~.
\end{equation}


\section{Numeric Model}
\label{appendix:num}
 For a time-dependent oscillator coupled to $N$ bath oscillators with an identical mass $m$, the composite Hamiltonian has the form:
\begin{multline}
\hat H=\f{\hat P^{2}}{2m}+\f 12m\omega^{2}\b t \hat Q^{2}\\+\sum_{i=1}^{N}\bigg(\f{p_{i}^{2}}{2m}+\f 12m\omega_{i}^{2} \hat q_{i}^{2}\bigg)+\hat Q\sum_i^{N} g_{k}\hat{q}_{k}
\end{multline}
where $\hat{p}_i$, $\hat{q}_i$ and  $\omega_i$ are momentum position and frequency of the $i$'th bath oscillator. 
The Heisenberg equations of motion are written in a vector-matrix form. For the vector $\v v$ the set of coupled differential equations are given by $\dot{\v v} \b t= {\cal{M}} \v v \b t$, where,
 \begin{multline}
      \v v =\big\{ \hat{Q}^2,\hat{P}^2,\f{\hat{Q}\hat{P}+\hat{P}\hat{Q}}{2},\hat{Q}\hat{q}_{1},\hat{Q}\hat{p}_{1},\hat{P}\hat{q}_{1},\\\hat{P}\hat{p}_{1},\hat{q}_{1}^{2}
     ,\hat{p}_{1}^{2},\f{\hat{q}_{1}\hat{p}_{1}+\hat{p}_{1}\hat{q}_{1}}{2},...\big\}^T~~,
 \label{eq:vector}
 \end{multline}
$\cal{M}=$
\begin{multline}
\tiny{
\left[{ \begin{array}{ccccccccccc}
0 & 0 & \f 2m & 0 & 0 & 0 & 0 & 0 & 0 & 0 & \cdots\\
0 & 0 & -2m\omega^{2} & 0 & 0 & -2g_{1} & 0 & 0 & 0 & 0\\
-m\omega^{2} & \f 1m & 0 & -g_{1} & 0 & 0 & 0 & 0 & 0 & 0\\
0 & 0 & 0 & 0 & \f 1m & \f 1m & 0 & 0 & 0 & 0\\
-g_{i} & 0 & 0 & -m\omega_{1}^{2} & 0 & 0 & \f 1m & 0 & 0 & 0\\
0 & 0 & 0 & -m\omega^{2} & 0 & 0 & \f 1m & -g_{1} & 0 & 0\\
0 & 0 & -g_{1} & 0 & -m\omega^{2} & -m\omega_{1}^{2} & 0 & 0 & 0 & -g_{1}\\
0 & 0 & 0 & 0 & 0 & 0 & 0 & 0 & 0 & \f 2m\\
0 & 0 & 0 & 0 & -2g_{1} & 0 & 0 & 0 & 0 & -2m\omega_{1}^{2}\\
0 & 0 & 0 & -g_{1} & 0 & 0 & 0 & -m\omega_{1}^{2} & \f 1m & 0\\
\vdots &  &  &  &  &  &  &  &  &  & \ddots
\end{array}}\right]}
\end{multline}

 The number of bath oscillators used to simulate the bath was $N=10^3$, which translates to $\sim 7\cdot10^{3}$ degrees of freedom (defining generalized Gaussian states, equivalent to the covariance matrix). This leads to a set of $\sim 7\cdot10^{3}$ coupled differential equations for the expectation values of the operators of $\v v\b t$, which describe the dynamics of the composite system. The set of differential equations are solved numerically using the Runge-Kutta-Fehlberg method with a time step $t_{step}=5\cdot10^{-4}$ (atomic units). 
  
\subsection{Numerical values}
\label{num:values}
The table is given in atomic units.
\begin{table}
\caption{Numerical values}
\begin{center}
\begin{tabular}{|l||c|}
\hline
\label{fig:constituents}
$\mu$ & $-10^{-5}$ \\ 
\hline
$\omega\b 0$ & $40$ \\ 
\hline
$\mean{\hat{H}\b 0}$ & $1.0375\cdot10^3$ \\ 
\hline
$\mean{\hat{L}\b 0}$ & $-5.625\cdot10^2$ \\ 
\hline
$\mean{\hat{C}\b 0}$ & $6\cdot 10^2$  \\ 
\hline
 Bath's spectral width ${\Delta \nu}$ & $\sb{0.6,1000}$  \\ 
 \hline
  Number of oscillators & $10^3$  \\ 
 \hline
 Oscillator mass ${m}$ & $2$    \\
 \hline
 Time step & $5\cdot10^{-4}$ \\
 \hline
 Coupling strength $g$ & $2.5\cdot 10^{-2}$   \\
  \hline
  $T_{bath}$ & $4$ \\ 
\hline
\end{tabular}
\end{center}
\end{table}


 \end{document}